\pdfoutput=1

\documentclass[pdflatex,sn-mathphys-num]{sn-jnl}% Math and Physical Sciences Numbered Reference Style 
\usepackage{graphicx}%
\DeclareGraphicsExtensions{.pdf,.png,.jpg,.jpeg}
\usepackage{multirow}%
\usepackage{amsmath,amssymb,amsfonts}%
\usepackage{amsthm}%
\usepackage{mathrsfs}%
\usepackage[title]{appendix}%
\usepackage{xcolor}%
\usepackage{textcomp}%
\usepackage{manyfoot}%
\usepackage{booktabs}%
\usepackage{algorithm}%
\usepackage{algorithmicx}%
\usepackage{algpseudocode}%
\usepackage{listings}%
\usepackage{braket}
\theoremstyle{thmstyleone}%
\theoremstyle{thmstyletwo}%

\theoremstyle{thmstylethree}%

\begin{document}

\title[Article Title]{Chip-to-chip hyperentanglement distribution and entanglement purification using silicon integrated photonics}

\author[1]{\fnm{Yonghe} \sur{Yu}}\email{yonyu@dtu.dk}


%\author[1]{\fnm{Cagin} \sur{Ekici}}\email{caek@dtu.dk}

\author[1]{\fnm{Mujtaba} \sur{Zahidy}}\email{muzah@dtu.dk}
\author[2]{\fnm{Siyan} \sur{Zhou}}\email{siyan.zhou@siphotonic.com}
\author[1]{\fnm{Caterina} \sur{Viligar}}\email{catvi@dtu.dk}
\author[1]{\fnm{Karsten} \sur{Rottwitt}}\email{karo@dtu.dk}
\author[1]{\fnm{Leif} \sur{Katsuo Oxenløwe}}\email{lkox@dtu.dk}
\author*[1]{\fnm{Yunhong} \sur{Ding}}\email{yudin@dtu.dk}

\affil[1]{\orgdiv{Department of Electrical and Photonics Engineering}, \orgname{Denmark Technical University}, \orgaddress{\street{Ørsteds Plads}, \city{Lyngby}, \postcode{2800}, \state{Hovedstaden}, \country{Denmark}}}

\affil[2]{\orgname{SiPhotonIC ApS}, \orgaddress{\city{Virum}, \postcode{2830}, \state{Hovedstaden}, \country{Denmark}}}


\abstract{Quantum repeaters are employed in quantum communication to overcome the long-distance transmission loss of quantum states. The quantum repeater is based on various key technologies, including quantum entanglement swapping, quantum memory, and entanglement purification. In particular, quantum purification can distil high-quality entanglement from the degraded entangled states which is propagating through noisy quantum communication channels. %To meet the requirements of scalability and cost-effectiveness of the next-generation quantum repeater, developing entanglement purification by integrated photonics is inevitable. 
Although previous reports have demonstrated on-chip entanglement swapping and teleportation through the less-noisy channel, current entanglement purification experiments still rely on off-chip discrete devices, 
%due to the absence of the on-chip genuine controlled NOT (CNOT) gates and on-chip purification protocols,
leading to limitations on scalability, stability and controllability. In this paper, for the first time, we demonstrated chip-to-chip hyperentanglment distribution and quantum entanglement purification based on integrated silicon chips. Path-encoded high-dimensional entangled photon pairs are produced on the chip, converted to fibre-based polarization-spatial hyperentanglement by grating couplers, distributed to the receiver silicon chip, and finally purified by consuming the spatial degree of freedom.
%Our purification process is free of post-selection and heralding, and thus, can be seamlessly integrated with other on-chip quantum information processing schemes that rely on entanglement distribution, including quantum communication and distributed quantum computing. 
Our purification scheme by integrated photonics finished the last puzzle of on-chip quantum repeater, which will promote the realization of the quantum repeater based on integrated photonics.
}
%这里有一种主打的是工业生产的意思，能不能把它的科学性给写出来？




\maketitle

\section{Introduction}\label{sec1}
Recent years, due to the rapid development of quantum communication\cite{lo2014secure,bouwmeester1997experimental} and distributed quantum computing\cite{ladd2010quantum,yimsiriwattana2004distributed}, quantum repeaters\cite{briegel1998quantum,sangouard2011quantum,pan2019qure,azuma2023quantum} attracted wide public interest because it promises the faithful transmission of the quantum state through long-distance quantum channel.
According to its protocol, the quantum repeater consists of three technologies\cite{briegel1998quantum,sangouard2011quantum,pan2019qure}: quantum entanglement swapping\cite{swapping11993,swapping21998,swapping32001}, quantum memory\cite{memory12000,memory22004,memory32009}, and entanglement purification\cite{puri41996,puri12001,puri22003,puri32021}. Since the essential purpose of the quantum repeater is to establish high-quality quantum entanglement between two distant communication ends\cite{puriandrepea1999}, entanglement purification is crucially important because it can distil high-quality entanglement from the degraded entangled states after propagating through the noisy quantum communication channel. 
In previous research \cite{puri41996,puriandrepea1999,puri12001,puripan32002,puri22003,puriraniner2006,puriZwerger2013,purichen2017,purikalb2017,puriwang2018,puriKrastanov2019,puririera2021,puri32021}, entanglement purification has been studied theoretically and experimentally using discrete bulk crystals and fiber-based devices\cite{puri12001,puri22003,puri32021}. Among them, deterministic entanglement purification based on hyperentanglement stands out because of its one-step scheme and high purification efficiency \cite{puri32021}.

To build a global quantum network with quantum repeaters, both cost and scalability must be considered. Free-space optical systems often suffer from high cost and alignment complexity, which severely limit their scalability \cite{miller2009device,zhu2021integrated}. In contrast, integrated photonics can integrate a lot of optical device on a single chip, which has emerged as a promising platform for implementing quantum protocols due to its high electronic and photonic integration, scalability, stability, and compatibility with complementary metal-oxide-semiconductor (CMOS) technology \cite{pelucchi2022potential}. Therefore, compared with discrete devices, directly producing independent bulk optical devices for global deployment of quantum repeaters offers no advantage. 
Therefore, compared with discrete devices, integrated photonics is more suitable for the industrial production of quantum repeaters for global deployment.
In fact, since entanglement swapping \cite{samara2021entanglement,llewellyn2020chipswapping} and quantum memory \cite{zhong2017nanophotonic,liu2020demand,cagin1} have already been realized on chip, the absence of on-chip quantum entanglement purification remains the last obstacle to the industrial production and world-wide deployment of quantum repeaters.

In this work, we demonstrate hyperentanglement-assisted entanglement purification \cite{puri32021} using silicon integrated photonics. To generate hyperentanglement in fibers, we first create the path-encoded high-dimensional bit in the silicon waveguides and then convert it into polarization-spatial hyperentanglement in fiber through 2D grating couplers (GC). After transmission through 3 m of single-mode fibers, the photons are received by another silicon chip, where entanglement purification is implemented with reconfigurable photonic circuits. A chip-to-chip optical phase-locked loop is proposed here to stabilize the phase difference between path-encoded bit 0 and bit 1.  
The performance of the system is verified under bit-flip (BF) and phase-flip (PF) noise. With our on-chip circuit, when a 20\% BF error rate is applied, the entanglement fidelity increases from 0.738 to 0.848 after purification, while the value of the Clauser-Horne-Shimony-Holt (CHSH) inequality \cite{CHSH1} rises from 1.898 to 2.195.

\section{Experimental setup}\label{sec2}
\begin{figure}[t]
		\mbox{%
			\includegraphics[width=13.1cm]{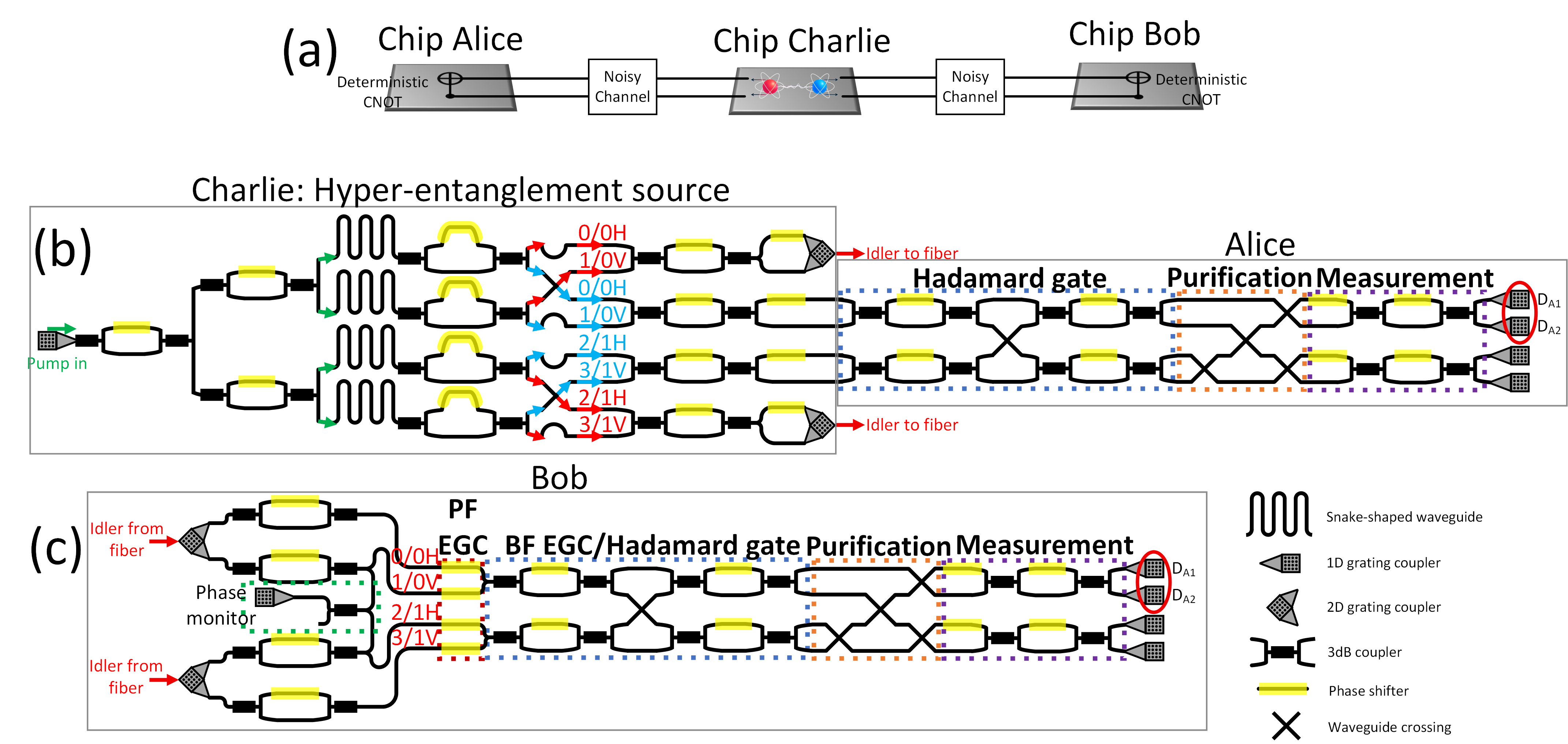}%
		}\caption{\label{fig1} The chip-based entanglement purification schematic and chip layout. 
        (a) Schematic of chip-based entanglement purification using hyperentanglement. The hyperentanglement source chip (Charlie) generates polarization-spatial-mode entangled photons, which are distributed to chip Alice and chip Bob. On chips Alice and Bob, the deterministic CNOT operations are implemented. After passing through the purification circuits on chips Alice and Bob, the entanglement fidelity of the polarization qubit is expected to be improved.
        %下面这个(b)感觉不可避免的会和正文重复
        (b) Schematic of chip Alice/Charlie. The integrated circuit on this silicon chip is responsible for entanglement generation, purification of signal photons (blue), and waveguide-to-fiber coupling of the idler photons (red). The pump pulses (green) are distributed into four snake-shaped waveguides where spontaneous four-wave mixing (SFWM) occurs. Only one photon pair is generated in four waveguides, and then demultiplexed by asymmetric MZIs composed of multimode interferometers (MMIs) and phase shifters. The `PF EGC’ represents the phase-flip error generation circuit, and the `BF EGC’ represents the bit-flip error generation circuit. The purification section consists of four reconfigurable waveguide crossings. For the idler photons, they are converted into polarization-spatial-mode entangled photons in fibers by the 2D grating coupler (2D GC).
        (c) Schematic of chip Bob. The pump and idler photons are coupled into the chip through the 2D GC. With the 2D GC, the polarization-spatial-mode encoded photons in fiber are converted back into path-encoded photons on chip. The first four MZIs act as mode attenuators to compensate for the different losses of the four optical paths and to split part of the pump into the central MMI of the phase monitor circuit. Here, about 11\% of the pump from the two fibers is directed into the phase monitor to interfere, converting the relative phase difference between the two fibers into a power difference at the outputs of this MMI.
        %(c), The rule of 2D GC on converting the qudit in waveguides to polarization-spatial qubit in fibers. (d), Schematic of our purification experiment and chip-to-chip phase locking. There are two single-mode fibers connecting the chip A and chip B. The PLL board reads the power from the `Pump couple out' in (b) and controls the phase shifter on the second fiber to finish the phase locking.
        }
	\end{figure}

%我想先介绍图一，再介绍图二，但是我又想先放图一，再放图二，有些矛盾
%目前的想法是先介绍图1a,说我们有两个片子，然后各是什么功能，purification是怎么运作的。然后说大致可以看一下图1b和图1c，之后介绍图2，介绍一下实验用的各种参数，然后就开始正式介绍
An ideal scheme for chip-to-chip hyperentanglement distribution and purification includes three chips, as shown in Fig.~\ref{fig1}(a). Chip Charlie performs hyperentanglement generation, and chips Alice and Bob receive the entanglement and carry out purification. In a field experiment through long distance fibers, the three chips are located at remote sites and connected by long fibers. The received photons may lose entanglement because of fiber vibration and background noise, and the entanglement fidelity should be improved after the purification on chip Alice and chip Bob. The purification operation here is based on the deterministic CNOT gate, where the polarization serves as the target qubit and the spatial-mode serves as the control qubit.

\begin{figure}[t]
		\mbox{%
			\includegraphics[width=13.1cm]{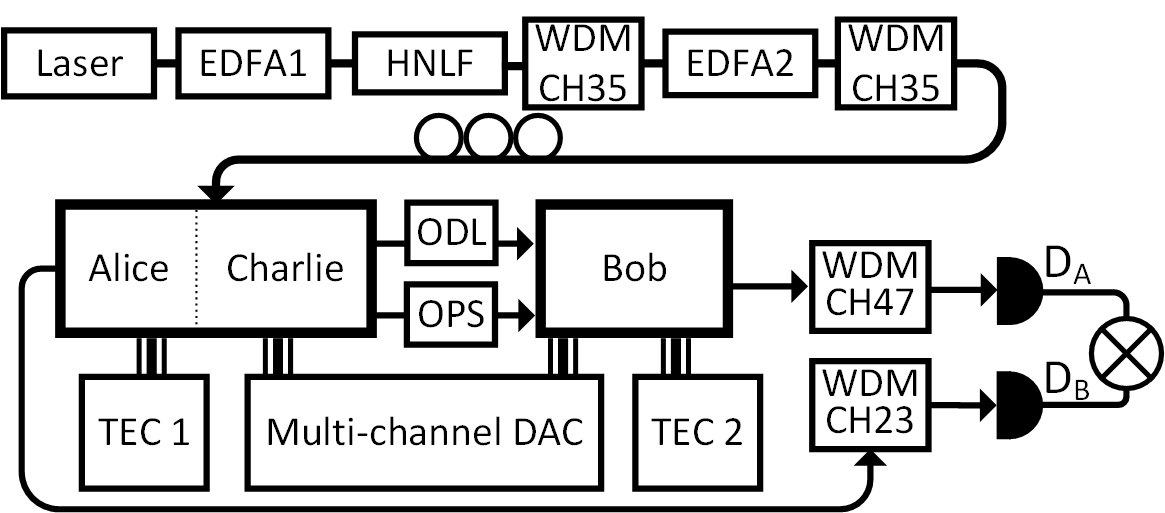}%
		}\caption{\label{fig2} The experimental setup for chip-to-chip hyperentanglement distribution and chip-based entanglement purification.
        Both silicon chips (chip Alice/Charlie and chip Bob) are optically and electrically packaged. Optical signals are coupled through single-mode fiber arrays to the GCs on both chips, and electrical control is provided via electrical cables. All cables are connected to a 96-channel digital-to-analog converter (DAC), which enables independent phase tuning of the heaters and reconfiguration of all the MZIs on both silicon chips. Each chip is also packaged with a thermoresistor and a thermoelectric cooler (TEC), allowing temperature stabilization through a TEC controller. The detection efficiency of the superconducting nanowire single-photon detectors (SNSPDs) is about 90\%, and the dark count rate and background noise are around 200 Hz.%是不是这里再写详细些
        }
	\end{figure}

In this work, to verify the feasibility of our scheme, the circuits for chip Alice and chip Charlie are prepared on one silicon chip (Fig.~\ref{fig1}(b)) and connected by waveguides. The circuit for chip Bob is fabricated on another silicon chip (Fig.~\ref{fig1}(c)), which is connected to chip Alice/Charlie through two single mode fibers for hyperentanglement transmission. 
The full experimental setup is shown in Fig.~\ref{fig2}. A 10 GHz repetition rate (measured as 9.95 GHz) pulsed laser serves as the pump source. The laser is first amplified by a pre-amplifier (EDFA1), and its spectrum is broadened by a highly nonlinear fiber (HNLF) to ensure spectral components around the central wavelength of 1549.32 nm, matching the passband of the 200-GHz wavelength-division multiplexing (WDM) filter. The selected light is further amplified by a high-power amplifier (EDFA2) and cleaned by a cascaded WDM filter at 1549.32 nm. After passing through a polarization controller (PC), the pump power before entering chip Alice/Charlie is 18.9 dBm. The pump is coupled into the 1D GC on chip Alice/Charlie using a single-mode fiber array, where the hyperentangled photons are generated in the Charlie part. Since the hyperentanglement is transmitted to chip Bob via two single mode fibers, the phase locking technology is employed to stabilize the relative phase that would otherwise drift randomly with time. On one fiber, a free space optical delay line (ODL) is inserted to align the temporal overlap of the two spatial modes. On the other fiber, a phase shifter (PS) controlled by the phase-locked loop controller (PLL controller) actively stabilizes the phase between the two fibers. After purification by the on-chip circuits, the photons are coupled out through a 1D grating coupler, and residual pump photons are suppressed by 100-GHz WDM filters at 1539.77 nm and 1558.89 nm. Finally, the photons are detected by superconducting nanowire single-photon detectors (SNSPDs), and coincidence events are recorded by a time tagger (TT).
%%等一下pferror的时候可能需要提一嘴，就是值得一提的是由光纤不稳定造成的phase drifting本身是可以被pfl的线路给纯化的，但是我们为了定量的分析pfl，前面还是加了相位补偿装置来给他补偿掉

%%下面这些我是这么个打算的，先写hyper entanglement generation,就是平均分光，然后2gc是怎么转化成 hyper entanglement。可以考虑一下要不要在图一加一下光纤的连接方式。然后写hyper entanglement transmission这中间也要写一下光纤我们是怎么做PLL的，包括前面那个mode attenuator怎么做的补偿。这之后我们可以介绍一下是怎么使用集成光子学线路模拟的bf pf，然后介绍一下 entanglement purification以及产生的结果

The entanglement is generated in the Charlie part of chip Alice/Charlie, as shown in Fig.~\ref{fig1}(b). To prepare the desired entangled state, three MZIs are used to evenly distribute the pump light from the 1D GC across four 1.5 cm snake-shaped waveguides. The pump intensity from the 1D GC is kept sufficiently low to ensure a small squeezing parameter (see Supplemental Material) and to suppress multi-photon pair emission, so that essentially only one or zero signal-idler photon pairs are generated in the four waveguides for each pump pulse. After the asymmetric Mach-Zehnder interferometers (AMZIs) and waveguide crossings, the signal-idler photon pair is high-dimensional path-entangled as  
$\ket{\phi}=\tfrac{1}{2}\left(\lvert 00 \rangle + \lvert 11 \rangle + \lvert 22 \rangle + \lvert 33 \rangle \right)$.  
Here, the numbers $0$-$3$ denote the waveguide path modes occupied by the signal or idler photon.

%这段是有用的，别删了  The free spectral range (FSR) of the AMZI is 322.7 GHz, which separates the signal photon at 1539.77 nm and the idler photon at 1558.98 nm, corresponding to a frequency spacing of 2400 GHz or 7.5 FSR. To achieve high spectral purity, a 200 GHz WDM filter with a bandwidth of 0.5 nm is used for the pump, and a 100 GHz WDM filter with a bandwidth of 0.22 nm is used for the signal and idler. To characterize the spectral purity of the photon pairs, we performed an unheralded Hanbury Brown-Twiss measurement of the idler photons, obtaining an unheralded $g^2(0)$ close to 2. From this, the spectral purity is calculated to be 94\% \cite{christ2011probing}. The measured squeezing parameter is 0.04, corresponding to a multi-photon fraction that is 25 times lower than the single-photon fraction.

\begin{figure}[t]
		\mbox{%
			\includegraphics[width=13.1cm]{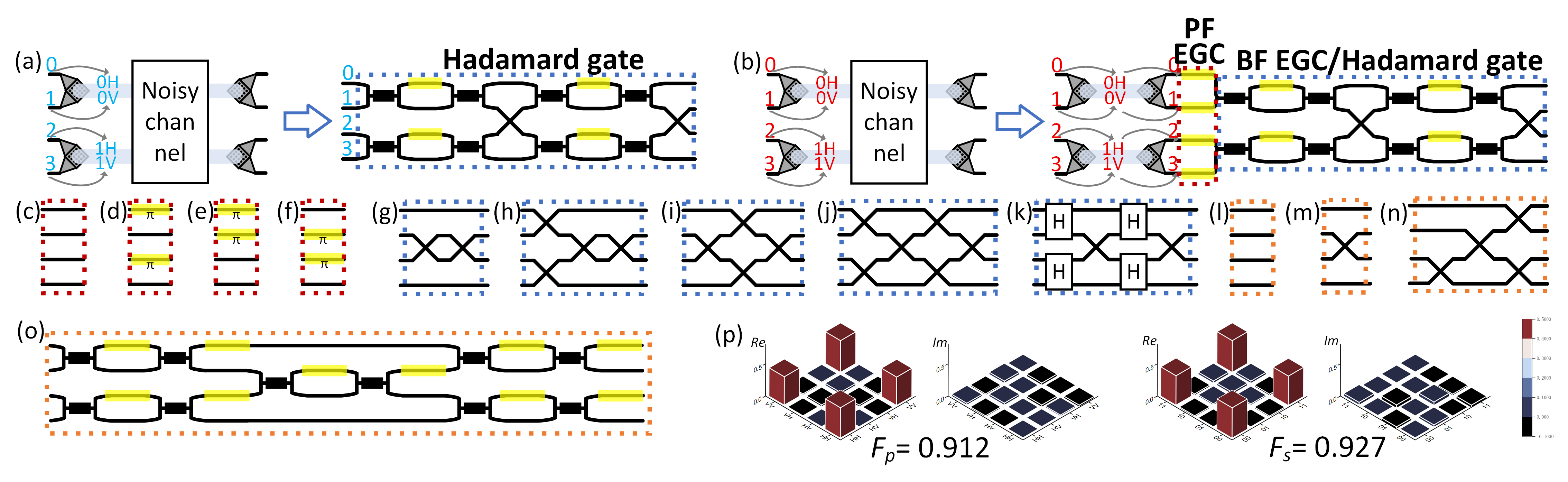}%
		}\caption{\label{fig3} Reconfigurable integrated circuits for noise simulation and purification demonstration in this work. (a) Hadamard gate circuit for the signal photon. (b) Phase-flip error generation circuit (PF EGC) and bit-flip error generation circuit (BF EGC) for the idler photon. ODL, PS, and on-chip mode attenuators are omitted in this figure for clarity.  
        (c)\textasciitilde(f) Reconfigured PF EGC for the cases of no error (c), PF error on the polarization degree of freedom (DOF) (d), PF error on the spatial-mode DOF (e), and PF errors on both degrees of freedom (f).  
        (g)\textasciitilde(j) Reconfigured BF EGC for the cases of no error (g), BF error on the polarization DOF (h), BF error on the spatial-mode DOF (i), and BF errors on both DOFs (j).  
        (k) Reconfigured Hadamard gate circuit used for PF error purification.  
        (l)\textasciitilde(n) Purification demonstration circuits for purification off with polarization QST (l), purification off with spatial QST (m), and purification on with polarization QST (n). (o) Original purification demonstration circuit constructed from five MZIs.  
        (p) Measured density matrices of polarization and spatial-mode qubits for the distributed hyperentangled states.
        }
	\end{figure}

%写下面这一段的时候，我在想是不是加一个类似之前ECOC里面的光纤连接图啊，问问丁老师吧
The path-encoded idler photon in the signal-idler entangled pair is converted into a polarization-spatial encoded qubit in fiber by the 2D GC. As marked in Fig.~\ref{fig3}(b), the path-encoded idler states $\ket{0}, \ket{1}, \ket{2}, \ket{3}$ are mapped to fiber-based polarization-spatial-mode encoded states $\ket{0H}, \ket{0V}, \ket{1H}, \ket{1V}$, respectively. 
If the 2D GC is also applied to the signal photons and the same mapping holds, the entangled state becomes a hyperentangled state $\ket{\phi} = \ket{\phi^+} \otimes \ket{\Phi^+} = \tfrac{1}{2}\left(\ket{00} + \ket{11}\right)\left(\ket{HH} + \ket{VV}\right)$. Here, $\ket{\phi^+}$ denotes one of the spatial-mode Bell states, and $\ket{\Phi^+}$ denotes one of the polarization Bell states.

%In two independent single mode fibers, even with shortl ength (e.g. 3m for our case), there could stil be random relative phase between tehm casued by temperateure or fibermove. in fact, this cause a PF error on spatial bit, which casue $\ket{\phi} = \ket{\phi^+} \otimes \ket{\Phi^+}$ goes to $\ket{\phi} = \ket{\phi^-} \otimes \ket{\Phi^+}$. 
Although our setup can correct PF errors caused by phase instability between the two single-mode fibers, to quantitatively demonstrate purification it is still necessary to keep the relative phase between the two fibers, which encode the spatial qubits 0 and 1, as stable as possible. The optical phase-locked loop (PLL) system implemented in this work consists of three parts: the on-chip phase monitor shown in Fig.~\ref{fig1}(c), the PLL controller, and a fiber-based phase shifter (PS) shown in Fig.~\ref{fig2}. On chip Bob, as shown in Fig.~\ref{fig1}(c), about 11\% of the pump light is tapped by the mode attenuator and interfered at the MMI of the phase monitor circuit. The output power from this MMI is coupled out through the 1D GC and sent to the PLL controller. The PLL controller reads the phase information from the detected power and drives the PS to compensate the phase difference. By setting the proportional-integral-derivative (PID) parameters of the PLL controller, we achieve stable phase locking with a power fluctuation of about $\pm 4.6\%$ (see Supplemental Material).

%相位锁定后，我们用量子态层析表征传输之后的超纠缠态。将error geneartion circuit与purification circuit 的mzi都设置为bar状态，再通过调节measurement circuit的mzi的heaters，我们可以对超纠缠中的偏振比特进行量子态层析。将purification circuit 的mzi重构为如图Fig. \ref{fig3}(m)所示，我们可以对超纠缠中的spatial mode比特进行量子态层析。如图Fig. \ref{fig3}(o)所示，经过2根3m单模光纤传输之后的光子偏振保真度为0.912\pm0.005，空间模保真度为0.927\pm0.004.保证度不完美，主要是来源于暗计数与sfwm的多光子分量。由于光栅耦合器造成了较大损耗（1dgc is 5.4dB, 2d gc 7dB, see Supplemental Material）, we only got a raw signal-idler coincident rate of about 10 Hz.此外，对于偏振比特而言，有限的2d gc消光比（14dB）可能造成偏振比特保真度下降。而对于空间模比特而言，锁相的相位浮动会造成保真度下降
After phase locking, we characterized the hyperentangled state after transmission through two 3-meter fibers using quantum state tomography (QST). By setting the MZIs in both the error generation circuit and the purification circuit to the bar state, and tuning the heaters of the MZIs in the measurement circuit, we performed QST on the polarization qubit of the hyperentangled state. The photons detected by the SNSPD are collected from the first two 1D GCs (red circles in Fig.~\ref{fig3}(b) and (c)).
By reconfiguring the MZIs in the purification circuit as shown in Fig.~\ref{fig3}(m), we performed tomography on the spatial-mode qubit (see Supplemental Material). As shown in Fig.~\ref{fig3}(o), after transmission the fibers, the polarization fidelity of the photons is measured to be $0.912 \pm 0.005$, and the spatial-mode fidelity is $0.927 \pm 0.004$. The imperfections in fidelity mainly arise from detector dark counts and multi-photon components generated by SFWM. Due to significant loss from the grating couplers (5.3 dB for the 1D GC and 5.9 dB for the 2D GC, see Supplemental Material), the raw signal-idler coincidence rate is about 10 Hz. In addition, for the polarization qubits, the limited extinction ratio of the 2D GC (16 dB) may reduce the fidelity, while for the spatial-mode qubits, residual phase fluctuations in the PLL contribute to fidelity degradation.

In this work, we use integrated circuits to simulate the PF and BF errors on the hyperentangled photons in fiber. For the idler photons, four heaters in the phase-flip error generation circuit (PF EGC), shown in the red dashed box of Fig.~\ref{fig3}(b), and four MZIs in the bit-flip error generation circuit (BF EGC), shown in the blue dashed box of Fig.~\ref{fig3}(b), are used to add additional flip errors. For the signal photons, as shown in Fig.~\ref{fig3}(a), we only implement Hadamard gate simulation circuits, since flip errors only need to be applied on one side.

\begin{figure}[t]
\centering
\includegraphics[width=\textwidth]{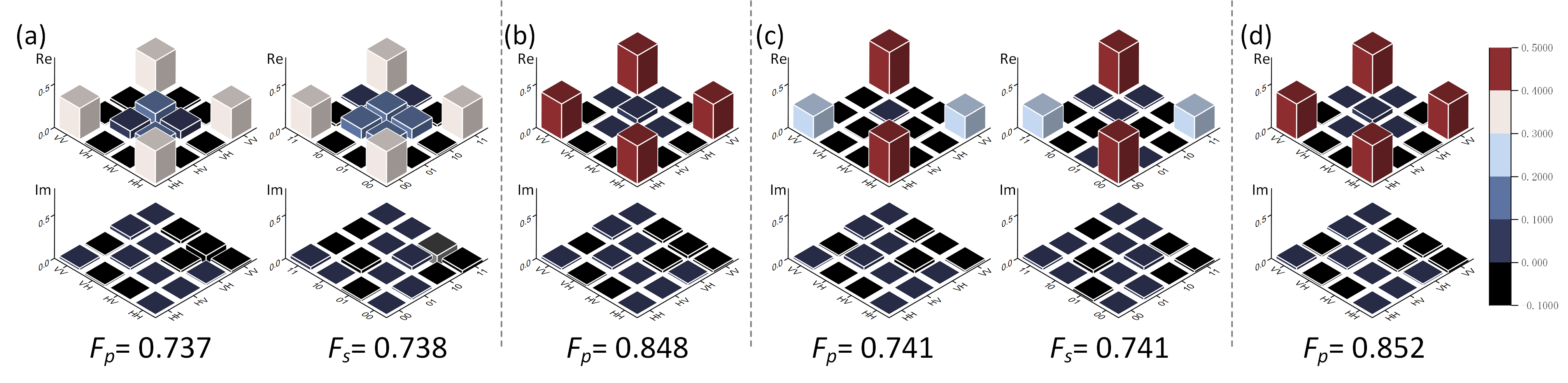}
\caption{QST results before and after purification. (a) Density matrices of the polarization qubit and spatial-mode qubit before purification with a 20\% BF error rate. (b) Density matrices of the polarization qubit after purification under a 20\% BF error rate. (c) Density matrices of the polarization qubit and spatial-mode qubit before purification with a 20\% PF error rate. (d) Density matrices of the polarization qubit after purification under a 20\% PF error rate.}\label{fig4}
\end{figure}

To introduce symmetrical BF noise on both the polarization and spatial-mode degrees of freedom (DOFs), we reconfigure the BF EGC into the circuits shown in Fig.~\ref{fig3}(g)\textasciitilde(j) with a time distribution \cite{puri32021} (see Supplemental Material). As noted earlier, the path qubit is mapped to the hyperentangled qubit in fiber, so swapping the photons in the waveguides is equivalent to inducing BF errors in the fiber. For example, by applying the circuit in Fig.~\ref{fig3}(h), the original state $\ket{\phi} = \ket{\phi^+} \otimes \ket{\Phi^+}$ is flipped to $\ket{\phi^+} \otimes \ket{\Psi^+} = \tfrac{1}{2}(\ket{00}+\ket{11})(\ket{HV}+\ket{VH})$. 
As shown in Fig.~\ref{fig4}(a), when a 20\% BF error rate is applied, the fidelity of the polarization qubit is $0.737 \pm 0.003$ with respect to the Bell state $\ket{\Phi^+} = \tfrac{1}{\sqrt{2}}(\ket{HH}+\ket{VV})$, and the fidelity of the spatial qubit is $0.738 \pm 0.003$ with respect to the Bell state $\ket{\phi^+} = \tfrac{1}{\sqrt{2}}(\ket{00}+\ket{11})$.

To demonstrate the PF noise before purification, we set the MZIs in the BF EGC to identity and introduced extra phase shifts with the heaters in the PF EGC. Still, to maintain symmetry on both the polarization and spatial-mode DOFs, the PF EGC is reconfigured into the circuits shown in Fig.~\ref{fig3}(c)\textasciitilde(f) with a certain time distribution. As shown in Fig.~\ref{fig4}(c), when a 20\% PF error rate is applied, the fidelity of the polarization qubit is $0.741 \pm 0.004$ with respect to the Bell state $\ket{\Phi^+} = \tfrac{1}{\sqrt{2}}(\ket{HH}+\ket{VV})$, and the fidelity of the spatial qubit is $0.741 \pm 0.003$ with respect to the Bell state $\ket{\phi^+} = \tfrac{1}{\sqrt{2}}(\ket{00}+\ket{11})$.

While noise simulation only needs to be implemented on either Alice or Bob, purification must be applied on both sides, as indicated by the orange dashed boxes in Fig.~\ref{fig1}(b) and Fig.~\ref{fig1}(c). The original purification circuit is build by five reconfigurable MZIs, as shown in Fig.~\ref{fig3}(o). When purification is enabled, one MZI is set to the identity state and the other four are set to the cross state, as illustrated in Fig.~\ref{fig3}(n). The purification circuit swaps photons between the waveguides on Alice’s and Bob’s chips, which corresponds to the following transformations in fiber-based hyperentangled state: $\ket{0H}\rightarrow\ket{0V}$, $\ket{0V}\rightarrow\ket{1V}$, $\ket{1H}\rightarrow\ket{1H}$, and $\ket{1V}\rightarrow\ket{0H}$. In the QST, as mentioned earlier, we take photons from waveguide $\ket{0}$ and waveguide $\ket{1}$ as the purified state and discard photons in waveguides $\ket{2}$ and $\ket{3}$. This is equivalent to discarding the fiber spatial-mode qubit \cite{puri32021}.
For the initial hyperentangled state $\ket{\phi}=\ket{\phi^+}\otimes\ket{\Phi^+}$, the purification operation leaves the state unchanged, and we obtain the entangled state $\ket{\Phi^+}$ from the first two waveguides (red circle in Fig.~\ref{fig1}). For an entangled state affected by BF errors in either the polarization or spatial dimension, one photon is preserved while the other is discarded, thereby causing no coincidence in the dataset and achieving purification (see Supplemental Material). 
Since the probability of BF errors occurring simultaneously on both the polarization and spatial-mode DOFs is much lower than the probability of errors on a single DOF, this scheme can significantly enhance entanglement fidelity. In this work, when a 20\% BF error rate is applied, as shown in Fig.~\ref{fig4}(b), the entanglement fidelity increases to $0.848\pm0.005$ after purification, representing an improvement of more than 11\%. This improvement is already very close to the 14\% theoretical limit of 20\% error rate (see Supplemental Material). We also measured the CHSH value using the measurement circuit, and it increased from $1.898\pm0.005$ to $2.195\pm0.007$ after purification under 20\% BF noise, thereby breaking the CHSH inequality.

To purify PF noise, it must first be converted into BF noise (see Supplemental Material). During the purification of PF noise, the four MZIs in the BF EGC are reconfigured into Hadamard gates, as shown in Fig.~\ref{fig3}(k). Because Hadamard gates are required on both sides, a similar circuit is added on the Alice side for the signal photons, as shown in Fig.~\ref{fig3}(a). After passing through the Hadamard gate circuit, the purification circuit is applied, and the entanglement fidelity of the polarization qubit under 20\% PF noise increases to $0.852 \pm 0.004$, as shown in Fig.~\ref{fig4}(d). Under the same 20\% PF noise, the CHSH value increases from $1.908 \pm 0.005$ to $2.200 \pm 0.007$ after purification, thereby violating the CHSH inequality.

\section{Discussion}\label{sec12}
%这个方案不能净化2bit的噪声，所以没法提升到满
%其实继承了cnot的功能
%丢掉的34路其实也能用
%与自由空间的方案相比，我们并不需要spatial与polar的跨维度交互，而是先通过2d gc统一转换为片上态，因此线路大大简化
The purification circuit proposed in this work inherently integrates the function of a CNOT gate, which is realized by a waveguide crossing that swaps the photons in waveguide modes $\ket{2}$ and $\ket{3}$. This operation is equivalent to flipping the polarization bit when the spatial bit is $\ket{1}$ for the fiber-based hyperentanglement. 

Similar to the free-space optical scheme \cite{puri32021}, our approach consumes one degree of freedom (DOF) of the entangled photons. In previous demonstrations, the discarded photons, which output from waveguide modes $\ket{2}$ and $\ket{3}$, could also be collected and were found to be purified as well (see Supplemental Material).

Compared with the hyperentanglement-based purification scheme using free-space optics \cite{puri32021}, our approach first converts the hyperentanglement into high-dimensional on-chip entanglement and performs all processing on chip, thereby avoiding CNOT operations across different DOFs. As a result, purification can be achieved with a simpler circuit design, rather than directly translating polarization beam splitters into on-chip fusion operations \cite{llewellyn2020chipswapping}.

Although the signal and idler photons are not both transmitted into fibers, meaning that a fully hyperentangled state is not realized, our approach of converting only the idler photons still demonstrates the feasibility of generating fiber-based hyperentanglement from a chip and chip-to-chip hyperentanglement distribution. To achieve a complete implementation, a PLL system would also need to be applied to the signal photons.

\section{Conclusion}\label{sec13}
%我们提出了一个集成光子学的方案来实现量子纠缠纯化，并且在硅基芯片实验演示。为了实现基于超纠缠的片上纯化方案，我们也一并证明了了使用芯片来生成光纤中的超纠缠态并实现超纠缠分发的可行性。拥有集成光子学带来的可扩展性，我们的纠缠纯化方案可以很容易的被集成到基于纠缠的其他量子通信或量子计算任务之前，并通过高效的纠缠纯化来提高这些任务的效果。而我们演示的芯片对芯片的超纠缠分发与芯片对芯片的锁相环本身也为量子通信提供了新思路。将我们的方案与基于集成光子学的纠缠交换与量子存储相结合，完全片上的量子中继器is ready to be achieved.我们硅光子学的工艺与cmos兼容，很适合用于工业生产，因此我们的工作可能导致量子中继器的大范围工业化应用。集成光子学是量子技术的平台的未来，我们的研究为将来基于集成光子学的量子中继器的发展指明了方向
We propose an integrated photonic scheme for entanglement purification and experimentally demonstrate it on silicon chips.
To realize an on-chip purification scheme based on hyperentanglement, we also demonstrated the feasibility of generating fiber-based hyperentanglement using integrated photonics chip and distributing it between chips. Benefiting from the scalability of integrated photonics, our entanglement purification scheme can be easily incorporated before other entanglement-based quantum communication \cite{ma2007quantum} or quantum computing \cite{eisert2021entangling} tasks, where a improved entanglement fidelity can further enhance their performance. Moreover, the demonstrated chip-to-chip hyperentanglement distribution and phase-locking provide new insights for quantum communication \cite{kim2021noise}. By integrating our circuit with on-chip entanglement swapping \cite{llewellyn2020chipswapping} and quantum memory \cite{zhong2017nanophotonic}, a fully chip-based quantum repeater is now ready to be realized. Since our silicon photonic platform is CMOS-compatible and suitable for industrial fabrication, this work may enable large-scale practical deployment of quantum repeaters. 
%最后这句话有必要加吗
Integrated photonics represents the future platform for quantum technologies, and our study points the way toward the development of integrated photonic quantum repeaters.

\backmatter

\bmhead{Data availability}

All the data that support the plots within this paper are available from the corresponding author upon reasonable request.

\bmhead{Code availability}

The code for processing the data in this work are available from the corresponding author upon reasonable request.

\bmhead{Acknowledgements}
We acknowledge funding from Villum Fonden Young Investigator project QUANPIC (Ref. 00025298) and Danish National Research Foundation Center of Excellence, SPOC (Ref. DNRF123).

\bmhead{Author contribution}
All authors contributed to the discussion and development of the project. Y.Y. conceived the idea and performed the theoretical analysis. Y.D., M.Z., and C.V. helped organize the experiment and discussed the idea. S.Z. and Y.D. fabricated the device. Y.Y. performed the experiment. K.R., L.K.O., and Y.D. managed the project. Y.Y. processed the data and prepared the manuscript.

% \section*{Declarations}
% Some journals require declarations to be submitted in a standardised format. 

% \begin{appendices}
% \section{Section title of first appendix}\label{secA1}
% An appendix contains supplementary information that is not an essential part of the text itself but which may be helpful in providing a more comprehensive understanding of the research problem or it is information that is too cumbersome to be included in the body of the paper.
% \end{appendices}


\bibliography{sn-bibliography}% common bib file


\begin{thebibliography}{49}
\ifx \bisbn   \undefined \def \bisbn  #1{ISBN #1}\fi
\ifx \binits  \undefined \def \binits#1{#1}\fi
\ifx \bauthor  \undefined \def \bauthor#1{#1}\fi
\ifx \batitle  \undefined \def \batitle#1{#1}\fi
\ifx \bjtitle  \undefined \def \bjtitle#1{#1}\fi
\ifx \bvolume  \undefined \def \bvolume#1{\textbf{#1}}\fi
\ifx \byear  \undefined \def \byear#1{#1}\fi
\ifx \bissue  \undefined \def \bissue#1{#1}\fi
\ifx \bfpage  \undefined \def \bfpage#1{#1}\fi
\ifx \blpage  \undefined \def \blpage #1{#1}\fi
\ifx \burl  \undefined \def \burl#1{\textsf{#1}}\fi
\ifx \doiurl  \undefined \def \doiurl#1{\url{https://doi.org/#1}}\fi
\ifx \betal  \undefined \def \betal{\textit{et al.}}\fi
\ifx \binstitute  \undefined \def \binstitute#1{#1}\fi
\ifx \binstitutionaled  \undefined \def \binstitutionaled#1{#1}\fi
\ifx \bctitle  \undefined \def \bctitle#1{#1}\fi
\ifx \beditor  \undefined \def \beditor#1{#1}\fi
\ifx \bpublisher  \undefined \def \bpublisher#1{#1}\fi
\ifx \bbtitle  \undefined \def \bbtitle#1{#1}\fi
\ifx \bedition  \undefined \def \bedition#1{#1}\fi
\ifx \bseriesno  \undefined \def \bseriesno#1{#1}\fi
\ifx \blocation  \undefined \def \blocation#1{#1}\fi
\ifx \bsertitle  \undefined \def \bsertitle#1{#1}\fi
\ifx \bsnm \undefined \def \bsnm#1{#1}\fi
\ifx \bsuffix \undefined \def \bsuffix#1{#1}\fi
\ifx \bparticle \undefined \def \bparticle#1{#1}\fi
\ifx \barticle \undefined \def \barticle#1{#1}\fi
\bibcommenthead
\ifx \bconfdate \undefined \def \bconfdate #1{#1}\fi
\ifx \botherref \undefined \def \botherref #1{#1}\fi
\ifx \url \undefined \def \url#1{\textsf{#1}}\fi
\ifx \bchapter \undefined \def \bchapter#1{#1}\fi
\ifx \bbook \undefined \def \bbook#1{#1}\fi
\ifx \bcomment \undefined \def \bcomment#1{#1}\fi
\ifx \oauthor \undefined \def \oauthor#1{#1}\fi
\ifx \citeauthoryear \undefined \def \citeauthoryear#1{#1}\fi
\ifx \endbibitem  \undefined \def \endbibitem {}\fi
\ifx \bconflocation  \undefined \def \bconflocation#1{#1}\fi
\ifx \arxivurl  \undefined \def \arxivurl#1{\textsf{#1}}\fi
\csname PreBibitemsHook\endcsname

%%% 1
\bibitem[\protect\citeauthoryear{Lo et~al.}{2014}]{lo2014secure}
\begin{barticle}
\bauthor{\bsnm{Lo}, \binits{H.-K.}},
\bauthor{\bsnm{Curty}, \binits{M.}},
\bauthor{\bsnm{Tamaki}, \binits{K.}}:
\batitle{Secure quantum key distribution}.
\bjtitle{Nature Photonics}
\bvolume{8}(\bissue{8}),
\bfpage{595}--\blpage{604}
(\byear{2014})
\end{barticle}
\endbibitem

%%% 2
\bibitem[\protect\citeauthoryear{Bouwmeester et~al.}{1997}]{bouwmeester1997experimental}
\begin{barticle}
\bauthor{\bsnm{Bouwmeester}, \binits{D.}},
\bauthor{\bsnm{Pan}, \binits{J.-W.}},
\bauthor{\bsnm{Mattle}, \binits{K.}},
\bauthor{\bsnm{Eibl}, \binits{M.}},
\bauthor{\bsnm{Weinfurter}, \binits{H.}},
\bauthor{\bsnm{Zeilinger}, \binits{A.}}:
\batitle{Experimental quantum teleportation}.
\bjtitle{Nature}
\bvolume{390}(\bissue{6660}),
\bfpage{575}--\blpage{579}
(\byear{1997})
\end{barticle}
\endbibitem

%%% 3
\bibitem[\protect\citeauthoryear{Ladd et~al.}{2010}]{ladd2010quantum}
\begin{barticle}
\bauthor{\bsnm{Ladd}, \binits{T.D.}},
\bauthor{\bsnm{Jelezko}, \binits{F.}},
\bauthor{\bsnm{Laflamme}, \binits{R.}},
\bauthor{\bsnm{Nakamura}, \binits{Y.}},
\bauthor{\bsnm{Monroe}, \binits{C.}},
\bauthor{\bsnm{O’Brien}, \binits{J.L.}}:
\batitle{Quantum computers}.
\bjtitle{nature}
\bvolume{464}(\bissue{7285}),
\bfpage{45}--\blpage{53}
(\byear{2010})
\end{barticle}
\endbibitem

%%% 4
\bibitem[\protect\citeauthoryear{Yimsiriwattana and Lomonaco~Jr}{2004}]{yimsiriwattana2004distributed}
\begin{bchapter}
\bauthor{\bsnm{Yimsiriwattana}, \binits{A.}},
\bauthor{\bsnm{Lomonaco~Jr}, \binits{S.J.}}:
\bctitle{Distributed quantum computing: A distributed shor algorithm}.
In: \bbtitle{Quantum Information and Computation II},
vol. \bseriesno{5436},
pp. \bfpage{360}--\blpage{372}
(\byear{2004}).
\bcomment{SPIE}
\end{bchapter}
\endbibitem

%%% 5
\bibitem[\protect\citeauthoryear{Briegel et~al.}{1998}]{briegel1998quantum}
\begin{barticle}
\bauthor{\bsnm{Briegel}, \binits{H.-J.}},
\bauthor{\bsnm{D{\"u}r}, \binits{W.}},
\bauthor{\bsnm{Cirac}, \binits{J.I.}},
\bauthor{\bsnm{Zoller}, \binits{P.}}:
\batitle{Quantum repeaters: the role of imperfect local operations in quantum communication}.
\bjtitle{Physical Review Letters}
\bvolume{81}(\bissue{26}),
\bfpage{5932}
(\byear{1998})
\end{barticle}
\endbibitem

%%% 6
\bibitem[\protect\citeauthoryear{Sangouard et~al.}{2011}]{sangouard2011quantum}
\begin{barticle}
\bauthor{\bsnm{Sangouard}, \binits{N.}},
\bauthor{\bsnm{Simon}, \binits{C.}},
\bauthor{\bsnm{De~Riedmatten}, \binits{H.}},
\bauthor{\bsnm{Gisin}, \binits{N.}}:
\batitle{Quantum repeaters based on atomic ensembles and linear optics}.
\bjtitle{Reviews of Modern Physics}
\bvolume{83}(\bissue{1}),
\bfpage{33}
(\byear{2011})
\end{barticle}
\endbibitem

%%% 7
\bibitem[\protect\citeauthoryear{Li et~al.}{2019}]{pan2019qure}
\begin{barticle}
\bauthor{\bsnm{Li}, \binits{Z.-D.}},
\bauthor{\bsnm{Zhang}, \binits{R.}},
\bauthor{\bsnm{Yin}, \binits{X.-F.}},
\bauthor{\bsnm{Liu}, \binits{L.-Z.}},
\bauthor{\bsnm{Hu}, \binits{Y.}},
\bauthor{\bsnm{Fang}, \binits{Y.-Q.}},
\bauthor{\bsnm{Fei}, \binits{Y.-Y.}},
\bauthor{\bsnm{Jiang}, \binits{X.}},
\bauthor{\bsnm{Zhang}, \binits{J.}},
\bauthor{\bsnm{Li}, \binits{L.}}, \betal:
\batitle{Experimental quantum repeater without quantum memory}.
\bjtitle{Nature photonics}
\bvolume{13}(\bissue{9}),
\bfpage{644}--\blpage{648}
(\byear{2019})
\end{barticle}
\endbibitem

%%% 8
\bibitem[\protect\citeauthoryear{Azuma et~al.}{2023}]{azuma2023quantum}
\begin{barticle}
\bauthor{\bsnm{Azuma}, \binits{K.}},
\bauthor{\bsnm{Economou}, \binits{S.E.}},
\bauthor{\bsnm{Elkouss}, \binits{D.}},
\bauthor{\bsnm{Hilaire}, \binits{P.}},
\bauthor{\bsnm{Jiang}, \binits{L.}},
\bauthor{\bsnm{Lo}, \binits{H.-K.}},
\bauthor{\bsnm{Tzitrin}, \binits{I.}}:
\batitle{Quantum repeaters: From quantum networks to the quantum internet}.
\bjtitle{Reviews of Modern Physics}
\bvolume{95}(\bissue{4}),
\bfpage{045006}
(\byear{2023})
\end{barticle}
\endbibitem

%%% 9
\bibitem[\protect\citeauthoryear{Zukowski et~al.}{1993}]{swapping11993}
\begin{botherref}
\oauthor{\bsnm{Zukowski}, \binits{M.}},
\oauthor{\bsnm{Zeilinger}, \binits{A.}},
\oauthor{\bsnm{Horne}, \binits{M.}},
\oauthor{\bsnm{Ekert}, \binits{A.}}:
" event-ready-detectors" bell experiment via entanglement swapping.
Physical Review Letters
\textbf{71}(26)
(1993)
\end{botherref}
\endbibitem

%%% 10
\bibitem[\protect\citeauthoryear{Pan et~al.}{1998}]{swapping21998}
\begin{barticle}
\bauthor{\bsnm{Pan}, \binits{J.-W.}},
\bauthor{\bsnm{Bouwmeester}, \binits{D.}},
\bauthor{\bsnm{Weinfurter}, \binits{H.}},
\bauthor{\bsnm{Zeilinger}, \binits{A.}}:
\batitle{Experimental entanglement swapping: entangling photons that never interacted}.
\bjtitle{Physical review letters}
\bvolume{80}(\bissue{18}),
\bfpage{3891}
(\byear{1998})
\end{barticle}
\endbibitem

%%% 11
\bibitem[\protect\citeauthoryear{Jennewein et~al.}{2001}]{swapping32001}
\begin{barticle}
\bauthor{\bsnm{Jennewein}, \binits{T.}},
\bauthor{\bsnm{Weihs}, \binits{G.}},
\bauthor{\bsnm{Pan}, \binits{J.-W.}},
\bauthor{\bsnm{Zeilinger}, \binits{A.}}:
\batitle{Experimental nonlocality proof of quantum teleportation and entanglement swapping}.
\bjtitle{Physical review letters}
\bvolume{88}(\bissue{1}),
\bfpage{017903}
(\byear{2001})
\end{barticle}
\endbibitem

%%% 12
\bibitem[\protect\citeauthoryear{Kozhekin et~al.}{2000}]{memory12000}
\begin{barticle}
\bauthor{\bsnm{Kozhekin}, \binits{A.}},
\bauthor{\bsnm{M{\o}lmer}, \binits{K.}},
\bauthor{\bsnm{Polzik}, \binits{E.}}:
\batitle{Quantum memory for light}.
\bjtitle{Physical Review A}
\bvolume{62}(\bissue{3}),
\bfpage{033809}
(\byear{2000})
\end{barticle}
\endbibitem

%%% 13
\bibitem[\protect\citeauthoryear{Julsgaard et~al.}{2004}]{memory22004}
\begin{barticle}
\bauthor{\bsnm{Julsgaard}, \binits{B.}},
\bauthor{\bsnm{Sherson}, \binits{J.}},
\bauthor{\bsnm{Cirac}, \binits{J.I.}},
\bauthor{\bsnm{Fiur{\'a}{\v{s}}ek}, \binits{J.}},
\bauthor{\bsnm{Polzik}, \binits{E.S.}}:
\batitle{Experimental demonstration of quantum memory for light}.
\bjtitle{Nature}
\bvolume{432}(\bissue{7016}),
\bfpage{482}--\blpage{486}
(\byear{2004})
\end{barticle}
\endbibitem

%%% 14
\bibitem[\protect\citeauthoryear{Lvovsky et~al.}{2009}]{memory32009}
\begin{barticle}
\bauthor{\bsnm{Lvovsky}, \binits{A.I.}},
\bauthor{\bsnm{Sanders}, \binits{B.C.}},
\bauthor{\bsnm{Tittel}, \binits{W.}}:
\batitle{Optical quantum memory}.
\bjtitle{Nature photonics}
\bvolume{3}(\bissue{12}),
\bfpage{706}--\blpage{714}
(\byear{2009})
\end{barticle}
\endbibitem

%%% 15
\bibitem[\protect\citeauthoryear{Bennett et~al.}{1996}]{puri41996}
\begin{barticle}
\bauthor{\bsnm{Bennett}, \binits{C.H.}},
\bauthor{\bsnm{Brassard}, \binits{G.}},
\bauthor{\bsnm{Popescu}, \binits{S.}},
\bauthor{\bsnm{Schumacher}, \binits{B.}},
\bauthor{\bsnm{Smolin}, \binits{J.A.}},
\bauthor{\bsnm{Wootters}, \binits{W.K.}}:
\batitle{Purification of noisy entanglement and faithful teleportation via noisy channels}.
\bjtitle{Physical review letters}
\bvolume{76}(\bissue{5}),
\bfpage{722}
(\byear{1996})
\end{barticle}
\endbibitem

%%% 16
\bibitem[\protect\citeauthoryear{Pan et~al.}{2001}]{puri12001}
\begin{barticle}
\bauthor{\bsnm{Pan}, \binits{J.-W.}},
\bauthor{\bsnm{Simon}, \binits{C.}},
\bauthor{\bsnm{Brukner}, \binits{{\v{C}}.}},
\bauthor{\bsnm{Zeilinger}, \binits{A.}}:
\batitle{Entanglement purification for quantum communication}.
\bjtitle{Nature}
\bvolume{410}(\bissue{6832}),
\bfpage{1067}--\blpage{1070}
(\byear{2001})
\end{barticle}
\endbibitem

%%% 17
\bibitem[\protect\citeauthoryear{Pan et~al.}{2003}]{puri22003}
\begin{barticle}
\bauthor{\bsnm{Pan}, \binits{J.-W.}},
\bauthor{\bsnm{Gasparoni}, \binits{S.}},
\bauthor{\bsnm{Ursin}, \binits{R.}},
\bauthor{\bsnm{Weihs}, \binits{G.}},
\bauthor{\bsnm{Zeilinger}, \binits{A.}}:
\batitle{Experimental entanglement purification of arbitrary unknown states}.
\bjtitle{Nature}
\bvolume{423}(\bissue{6938}),
\bfpage{417}--\blpage{422}
(\byear{2003})
\end{barticle}
\endbibitem

%%% 18
\bibitem[\protect\citeauthoryear{Hu et~al.}{2021}]{puri32021}
\begin{barticle}
\bauthor{\bsnm{Hu}, \binits{X.-M.}},
\bauthor{\bsnm{Huang}, \binits{C.-X.}},
\bauthor{\bsnm{Sheng}, \binits{Y.-B.}},
\bauthor{\bsnm{Zhou}, \binits{L.}},
\bauthor{\bsnm{Liu}, \binits{B.-H.}},
\bauthor{\bsnm{Guo}, \binits{Y.}},
\bauthor{\bsnm{Zhang}, \binits{C.}},
\bauthor{\bsnm{Xing}, \binits{W.-B.}},
\bauthor{\bsnm{Huang}, \binits{Y.-F.}},
\bauthor{\bsnm{Li}, \binits{C.-F.}}, \betal:
\batitle{Long-distance entanglement purification for quantum communication}.
\bjtitle{Physical review letters}
\bvolume{126}(\bissue{1}),
\bfpage{010503}
(\byear{2021})
\end{barticle}
\endbibitem

%%% 19
\bibitem[\protect\citeauthoryear{D{\"u}r et~al.}{1999}]{puriandrepea1999}
\begin{barticle}
\bauthor{\bsnm{D{\"u}r}, \binits{W.}},
\bauthor{\bsnm{Briegel}, \binits{H.-J.}},
\bauthor{\bsnm{Cirac}, \binits{J.I.}},
\bauthor{\bsnm{Zoller}, \binits{P.}}:
\batitle{Quantum repeaters based on entanglement purification}.
\bjtitle{Physical Review A}
\bvolume{59}(\bissue{1}),
\bfpage{169}
(\byear{1999})
\end{barticle}
\endbibitem

%%% 20
\bibitem[\protect\citeauthoryear{Simon and Pan}{2002}]{puripan32002}
\begin{barticle}
\bauthor{\bsnm{Simon}, \binits{C.}},
\bauthor{\bsnm{Pan}, \binits{J.-W.}}:
\batitle{Polarization entanglement purification using spatial entanglement}.
\bjtitle{Phys. Rev. Lett.}
\bvolume{89},
\bfpage{257901}
(\byear{2002})
\doiurl{10.1103/PhysRevLett.89.257901}
\end{barticle}
\endbibitem

%%% 21
\bibitem[\protect\citeauthoryear{Reichle et~al.}{2006}]{puriraniner2006}
\begin{barticle}
\bauthor{\bsnm{Reichle}, \binits{R.}},
\bauthor{\bsnm{Leibfried}, \binits{D.}},
\bauthor{\bsnm{Knill}, \binits{E.}},
\bauthor{\bsnm{Britton}, \binits{J.}},
\bauthor{\bsnm{Blakestad}, \binits{R.B.}},
\bauthor{\bsnm{Jost}, \binits{J.D.}},
\bauthor{\bsnm{Langer}, \binits{C.}},
\bauthor{\bsnm{Ozeri}, \binits{R.}},
\bauthor{\bsnm{Seidelin}, \binits{S.}},
\bauthor{\bsnm{Wineland}, \binits{D.J.}}:
\batitle{Experimental purification of two-atom entanglement}.
\bjtitle{Nature}
\bvolume{443}(\bissue{7113}),
\bfpage{838}--\blpage{841}
(\byear{2006})
\end{barticle}
\endbibitem

%%% 22
\bibitem[\protect\citeauthoryear{Zwerger et~al.}{2013}]{puriZwerger2013}
\begin{barticle}
\bauthor{\bsnm{Zwerger}, \binits{M.}},
\bauthor{\bsnm{Briegel}, \binits{H.J.}},
\bauthor{\bsnm{D\"ur}, \binits{W.}}:
\batitle{Universal and optimal error thresholds for measurement-based entanglement purification}.
\bjtitle{Phys. Rev. Lett.}
\bvolume{110},
\bfpage{260503}
(\byear{2013})
\doiurl{10.1103/PhysRevLett.110.260503}
\end{barticle}
\endbibitem

%%% 23
\bibitem[\protect\citeauthoryear{Chen et~al.}{2017}]{purichen2017}
\begin{barticle}
\bauthor{\bsnm{Chen}, \binits{L.-K.}},
\bauthor{\bsnm{Yong}, \binits{H.-L.}},
\bauthor{\bsnm{Xu}, \binits{P.}},
\bauthor{\bsnm{Yao}, \binits{X.-C.}},
\bauthor{\bsnm{Xiang}, \binits{T.}},
\bauthor{\bsnm{Li}, \binits{Z.-D.}},
\bauthor{\bsnm{Liu}, \binits{C.}},
\bauthor{\bsnm{Lu}, \binits{H.}},
\bauthor{\bsnm{Liu}, \binits{N.-L.}},
\bauthor{\bsnm{Li}, \binits{L.}}, \betal:
\batitle{Experimental nested purification for a linear optical quantum repeater}.
\bjtitle{Nature Photonics}
\bvolume{11}(\bissue{11}),
\bfpage{695}--\blpage{699}
(\byear{2017})
\end{barticle}
\endbibitem

%%% 24
\bibitem[\protect\citeauthoryear{Kalb et~al.}{2017}]{purikalb2017}
\begin{barticle}
\bauthor{\bsnm{Kalb}, \binits{N.}},
\bauthor{\bsnm{Reiserer}, \binits{A.A.}},
\bauthor{\bsnm{Humphreys}, \binits{P.C.}},
\bauthor{\bsnm{Bakermans}, \binits{J.J.W.}},
\bauthor{\bsnm{Kamerling}, \binits{S.J.}},
\bauthor{\bsnm{Nickerson}, \binits{N.H.}},
\bauthor{\bsnm{Benjamin}, \binits{S.C.}},
\bauthor{\bsnm{Twitchen}, \binits{D.J.}},
\bauthor{\bsnm{Markham}, \binits{M.}},
\bauthor{\bsnm{Hanson}, \binits{R.}}:
\batitle{Entanglement distillation between solid-state quantum network nodes}.
\bjtitle{Science}
\bvolume{356}(\bissue{6341}),
\bfpage{928}--\blpage{932}
(\byear{2017})
\doiurl{10.1126/science.aan0070}
\end{barticle}
\endbibitem

%%% 25
\bibitem[\protect\citeauthoryear{Wang et~al.}{2018}]{puriwang2018}
\begin{barticle}
\bauthor{\bsnm{Wang}, \binits{G.-Y.}},
\bauthor{\bsnm{Li}, \binits{T.}},
\bauthor{\bsnm{Ai}, \binits{Q.}},
\bauthor{\bsnm{Alsaedi}, \binits{A.}},
\bauthor{\bsnm{Hayat}, \binits{T.}},
\bauthor{\bsnm{Deng}, \binits{F.-G.}}:
\batitle{Faithful entanglement purification for high-capacity quantum communication with two-photon four-qubit systems}.
\bjtitle{Phys. Rev. Appl.}
\bvolume{10},
\bfpage{054058}
(\byear{2018})
\doiurl{10.1103/PhysRevApplied.10.054058}
\end{barticle}
\endbibitem

%%% 26
\bibitem[\protect\citeauthoryear{Krastanov et~al.}{2019}]{puriKrastanov2019}
\begin{barticle}
\bauthor{\bsnm{Krastanov}, \binits{S.}},
\bauthor{\bsnm{Albert}, \binits{V.V.}},
\bauthor{\bsnm{Jiang}, \binits{L.}}:
\batitle{Optimized {E}ntanglement {P}urification}.
\bjtitle{{Quantum}}
\bvolume{3},
\bfpage{123}
(\byear{2019})
\doiurl{10.22331/q-2019-02-18-123}
\end{barticle}
\endbibitem

%%% 27
\bibitem[\protect\citeauthoryear{Riera-S\`abat et~al.}{2021}]{puririera2021}
\begin{barticle}
\bauthor{\bsnm{Riera-S\`abat}, \binits{F.}},
\bauthor{\bsnm{Sekatski}, \binits{P.}},
\bauthor{\bsnm{Pirker}, \binits{A.}},
\bauthor{\bsnm{D\"ur}, \binits{W.}}:
\batitle{Entanglement-assisted entanglement purification}.
\bjtitle{Phys. Rev. Lett.}
\bvolume{127},
\bfpage{040502}
(\byear{2021})
\doiurl{10.1103/PhysRevLett.127.040502}
\end{barticle}
\endbibitem

%%% 28
\bibitem[\protect\citeauthoryear{Miller}{2009}]{miller2009device}
\begin{barticle}
\bauthor{\bsnm{Miller}, \binits{D.A.}}:
\batitle{Device requirements for optical interconnects to silicon chips}.
\bjtitle{Proceedings of the IEEE}
\bvolume{97}(\bissue{7}),
\bfpage{1166}--\blpage{1185}
(\byear{2009})
\end{barticle}
\endbibitem

%%% 29
\bibitem[\protect\citeauthoryear{Zhu et~al.}{2021}]{zhu2021integrated}
\begin{barticle}
\bauthor{\bsnm{Zhu}, \binits{D.}},
\bauthor{\bsnm{Shao}, \binits{L.}},
\bauthor{\bsnm{Yu}, \binits{M.}},
\bauthor{\bsnm{Cheng}, \binits{R.}},
\bauthor{\bsnm{Desiatov}, \binits{B.}},
\bauthor{\bsnm{Xin}, \binits{C.}},
\bauthor{\bsnm{Hu}, \binits{Y.}},
\bauthor{\bsnm{Holzgrafe}, \binits{J.}},
\bauthor{\bsnm{Ghosh}, \binits{S.}},
\bauthor{\bsnm{Shams-Ansari}, \binits{A.}}, \betal:
\batitle{Integrated photonics on thin-film lithium niobate}.
\bjtitle{Advances in Optics and Photonics}
\bvolume{13}(\bissue{2}),
\bfpage{242}--\blpage{352}
(\byear{2021})
\end{barticle}
\endbibitem

%%% 30
\bibitem[\protect\citeauthoryear{Pelucchi et~al.}{2022}]{pelucchi2022potential}
\begin{barticle}
\bauthor{\bsnm{Pelucchi}, \binits{E.}},
\bauthor{\bsnm{Fagas}, \binits{G.}},
\bauthor{\bsnm{Aharonovich}, \binits{I.}},
\bauthor{\bsnm{Englund}, \binits{D.}},
\bauthor{\bsnm{Figueroa}, \binits{E.}},
\bauthor{\bsnm{Gong}, \binits{Q.}},
\bauthor{\bsnm{Hannes}, \binits{H.}},
\bauthor{\bsnm{Liu}, \binits{J.}},
\bauthor{\bsnm{Lu}, \binits{C.-Y.}},
\bauthor{\bsnm{Matsuda}, \binits{N.}}, \betal:
\batitle{The potential and global outlook of integrated photonics for quantum technologies}.
\bjtitle{Nature Reviews Physics}
\bvolume{4}(\bissue{3}),
\bfpage{194}--\blpage{208}
(\byear{2022})
\end{barticle}
\endbibitem

%%% 31
\bibitem[\protect\citeauthoryear{Samara et~al.}{2021}]{samara2021entanglement}
\begin{barticle}
\bauthor{\bsnm{Samara}, \binits{F.}},
\bauthor{\bsnm{Maring}, \binits{N.}},
\bauthor{\bsnm{Martin}, \binits{A.}},
\bauthor{\bsnm{Raja}, \binits{A.S.}},
\bauthor{\bsnm{Kippenberg}, \binits{T.J.}},
\bauthor{\bsnm{Zbinden}, \binits{H.}},
\bauthor{\bsnm{Thew}, \binits{R.}}:
\batitle{Entanglement swapping between independent and asynchronous integrated photon-pair sources}.
\bjtitle{Quantum Science and Technology}
\bvolume{6}(\bissue{4}),
\bfpage{045024}
(\byear{2021})
\doiurl{10.1088/2058-9565/abf599}
\end{barticle}
\endbibitem

%%% 32
\bibitem[\protect\citeauthoryear{Llewellyn et~al.}{2020}]{llewellyn2020chipswapping}
\begin{barticle}
\bauthor{\bsnm{Llewellyn}, \binits{D.}},
\bauthor{\bsnm{Ding}, \binits{Y.}},
\bauthor{\bsnm{Faruque}, \binits{I.I.}},
\bauthor{\bsnm{Paesani}, \binits{S.}},
\bauthor{\bsnm{Bacco}, \binits{D.}},
\bauthor{\bsnm{Santagati}, \binits{R.}},
\bauthor{\bsnm{Qian}, \binits{Y.-J.}},
\bauthor{\bsnm{Li}, \binits{Y.}},
\bauthor{\bsnm{Xiao}, \binits{Y.-F.}},
\bauthor{\bsnm{Huber}, \binits{M.}}, \betal:
\batitle{Chip-to-chip quantum teleportation and multi-photon entanglement in silicon}.
\bjtitle{Nature Physics}
\bvolume{16}(\bissue{2}),
\bfpage{148}--\blpage{153}
(\byear{2020})
\end{barticle}
\endbibitem

%%% 33
\bibitem[\protect\citeauthoryear{Zhong et~al.}{2017}]{zhong2017nanophotonic}
\begin{barticle}
\bauthor{\bsnm{Zhong}, \binits{T.}},
\bauthor{\bsnm{Kindem}, \binits{J.M.}},
\bauthor{\bsnm{Bartholomew}, \binits{J.G.}},
\bauthor{\bsnm{Rochman}, \binits{J.}},
\bauthor{\bsnm{Craiciu}, \binits{I.}},
\bauthor{\bsnm{Miyazono}, \binits{E.}},
\bauthor{\bsnm{Bettinelli}, \binits{M.}},
\bauthor{\bsnm{Cavalli}, \binits{E.}},
\bauthor{\bsnm{Verma}, \binits{V.}},
\bauthor{\bsnm{Nam}, \binits{S.W.}}, \betal:
\batitle{Nanophotonic rare-earth quantum memory with optically controlled retrieval}.
\bjtitle{Science}
\bvolume{357}(\bissue{6358}),
\bfpage{1392}--\blpage{1395}
(\byear{2017})
\end{barticle}
\endbibitem

%%% 34
\bibitem[\protect\citeauthoryear{Liu et~al.}{2020}]{liu2020demand}
\begin{barticle}
\bauthor{\bsnm{Liu}, \binits{C.}},
\bauthor{\bsnm{Zhu}, \binits{T.-X.}},
\bauthor{\bsnm{Su}, \binits{M.-X.}},
\bauthor{\bsnm{Ma}, \binits{Y.-Z.}},
\bauthor{\bsnm{Zhou}, \binits{Z.-Q.}},
\bauthor{\bsnm{Li}, \binits{C.-F.}},
\bauthor{\bsnm{Guo}, \binits{G.-C.}}:
\batitle{On-demand quantum storage of photonic qubits in an on-chip waveguide}.
\bjtitle{Physical Review Letters}
\bvolume{125}(\bissue{26}),
\bfpage{260504}
(\byear{2020})
\end{barticle}
\endbibitem

%%% 35
\bibitem[\protect\citeauthoryear{Ekici et~al.}{2025}]{cagin1}
\begin{barticle}
\bauthor{\bsnm{Ekici}, \binits{{\c{C}}.}},
\bauthor{\bsnm{Yu}, \binits{Y.}},
\bauthor{\bsnm{Adcock}, \binits{J.C.}},
\bauthor{\bsnm{Muthali}, \binits{A.L.}},
\bauthor{\bsnm{Tan}, \binits{H.}},
\bauthor{\bsnm{Lin}, \binits{Z.}},
\bauthor{\bsnm{Li}, \binits{H.}},
\bauthor{\bsnm{Oxenl{\o}we}, \binits{L.K.}},
\bauthor{\bsnm{Cai}, \binits{X.}},
\bauthor{\bsnm{Ding}, \binits{Y.}}:
\batitle{High-resolution single photon level storage of telecom light based on thin film lithium niobate photonics}.
\bjtitle{Advanced Quantum Technologies}
\bvolume{8}(\bissue{2}),
\bfpage{2300195}
(\byear{2025})
\end{barticle}
\endbibitem

%%% 36
\bibitem[\protect\citeauthoryear{Aspect et~al.}{1982}]{CHSH1}
\begin{barticle}
\bauthor{\bsnm{Aspect}, \binits{A.}},
\bauthor{\bsnm{Grangier}, \binits{P.}},
\bauthor{\bsnm{Roger}, \binits{G.}}:
\batitle{Experimental realization of einstein-podolsky-rosen-bohm gedankenexperiment: A new violation of bell's inequalities}.
\bjtitle{Phys. Rev. Lett.}
\bvolume{49},
\bfpage{91}--\blpage{94}
(\byear{1982})
\doiurl{10.1103/PhysRevLett.49.91}
\end{barticle}
\endbibitem

%%% 37
\bibitem[\protect\citeauthoryear{Ma et~al.}{2007}]{ma2007quantum}
\begin{barticle}
\bauthor{\bsnm{Ma}, \binits{X.}},
\bauthor{\bsnm{Fung}, \binits{C.-H.F.}},
\bauthor{\bsnm{Lo}, \binits{H.-K.}}:
\batitle{Quantum key distribution with entangled photon sources}.
\bjtitle{Physical Review A—Atomic, Molecular, and Optical Physics}
\bvolume{76}(\bissue{1}),
\bfpage{012307}
(\byear{2007})
\end{barticle}
\endbibitem

%%% 38
\bibitem[\protect\citeauthoryear{Eisert}{2021}]{eisert2021entangling}
\begin{barticle}
\bauthor{\bsnm{Eisert}, \binits{J.}}:
\batitle{Entangling power and quantum circuit complexity}.
\bjtitle{Physical Review Letters}
\bvolume{127}(\bissue{2}),
\bfpage{020501}
(\byear{2021})
\end{barticle}
\endbibitem

%%% 39
\bibitem[\protect\citeauthoryear{Kim et~al.}{2021}]{kim2021noise}
\begin{barticle}
\bauthor{\bsnm{Kim}, \binits{J.-H.}},
\bauthor{\bsnm{Kim}, \binits{Y.}},
\bauthor{\bsnm{Im}, \binits{D.-G.}},
\bauthor{\bsnm{Lee}, \binits{C.-H.}},
\bauthor{\bsnm{Chae}, \binits{J.-W.}},
\bauthor{\bsnm{Scarcelli}, \binits{G.}},
\bauthor{\bsnm{Kim}, \binits{Y.-H.}}:
\batitle{Noise-resistant quantum communications using hyperentanglement}.
\bjtitle{Optica}
\bvolume{8}(\bissue{12}),
\bfpage{1524}--\blpage{1531}
(\byear{2021})
\end{barticle}
\endbibitem

%%% 40
\bibitem[\protect\citeauthoryear{Xue et~al.}{2019}]{xue2019two}
\begin{barticle}
\bauthor{\bsnm{Xue}, \binits{Y.}},
\bauthor{\bsnm{Chen}, \binits{H.}},
\bauthor{\bsnm{Bao}, \binits{Y.}},
\bauthor{\bsnm{Dong}, \binits{J.}},
\bauthor{\bsnm{Zhang}, \binits{X.}}:
\batitle{Two-dimensional silicon photonic grating coupler with low polarization-dependent loss and high tolerance}.
\bjtitle{Optics express}
\bvolume{27}(\bissue{16}),
\bfpage{22268}--\blpage{22274}
(\byear{2019})
\end{barticle}
\endbibitem

%%% 41
\bibitem[\protect\citeauthoryear{Silverstone et~al.}{2014}]{silverstone2014chip}
\begin{barticle}
\bauthor{\bsnm{Silverstone}, \binits{J.W.}},
\bauthor{\bsnm{Bonneau}, \binits{D.}},
\bauthor{\bsnm{Ohira}, \binits{K.}},
\bauthor{\bsnm{Suzuki}, \binits{N.}},
\bauthor{\bsnm{Yoshida}, \binits{H.}},
\bauthor{\bsnm{Iizuka}, \binits{N.}},
\bauthor{\bsnm{Ezaki}, \binits{M.}},
\bauthor{\bsnm{Natarajan}, \binits{C.M.}},
\bauthor{\bsnm{Tanner}, \binits{M.G.}},
\bauthor{\bsnm{Hadfield}, \binits{R.H.}}, \betal:
\batitle{On-chip quantum interference between silicon photon-pair sources}.
\bjtitle{Nature Photonics}
\bvolume{8}(\bissue{2}),
\bfpage{104}--\blpage{108}
(\byear{2014})
\end{barticle}
\endbibitem

%%% 42
\bibitem[\protect\citeauthoryear{Wang et~al.}{2018}]{wang2018multienta}
\begin{barticle}
\bauthor{\bsnm{Wang}, \binits{J.}},
\bauthor{\bsnm{Paesani}, \binits{S.}},
\bauthor{\bsnm{Ding}, \binits{Y.}},
\bauthor{\bsnm{Santagati}, \binits{R.}},
\bauthor{\bsnm{Skrzypczyk}, \binits{P.}},
\bauthor{\bsnm{Salavrakos}, \binits{A.}},
\bauthor{\bsnm{Tura}, \binits{J.}},
\bauthor{\bsnm{Augusiak}, \binits{R.}},
\bauthor{\bsnm{Man{\v{c}}inska}, \binits{L.}},
\bauthor{\bsnm{Bacco}, \binits{D.}}, \betal:
\batitle{Multidimensional quantum entanglement with large-scale integrated optics}.
\bjtitle{Science}
\bvolume{360},
\bfpage{285}--\blpage{291}
(\byear{2018})
\end{barticle}
\endbibitem

%%% 43
\bibitem[\protect\citeauthoryear{Adcock et~al.}{2019}]{adcock2019programmable}
\begin{barticle}
\bauthor{\bsnm{Adcock}, \binits{J.C.}},
\bauthor{\bsnm{Vigliar}, \binits{C.}},
\bauthor{\bsnm{Santagati}, \binits{R.}},
\bauthor{\bsnm{Silverstone}, \binits{J.W.}},
\bauthor{\bsnm{Thompson}, \binits{M.G.}}:
\batitle{Programmable four-photon graph states on a silicon chip}.
\bjtitle{Nature communications}
\bvolume{10}(\bissue{1}),
\bfpage{3528}
(\byear{2019})
\end{barticle}
\endbibitem

%%% 44
\bibitem[\protect\citeauthoryear{Llewellyn}{2020}]{danielthesis}
\begin{botherref}
\oauthor{\bsnm{Llewellyn}, \binits{D.M.}}:
Quantum information processing by programming optical nano-circuits in silicon.
PhD thesis,
University of Bristol
(2020)
\end{botherref}
\endbibitem

%%% 45
\bibitem[\protect\citeauthoryear{Christ et~al.}{2011}]{christ2011probing}
\begin{barticle}
\bauthor{\bsnm{Christ}, \binits{A.}},
\bauthor{\bsnm{Laiho}, \binits{K.}},
\bauthor{\bsnm{Eckstein}, \binits{A.}},
\bauthor{\bsnm{Cassemiro}, \binits{K.N.}},
\bauthor{\bsnm{Silberhorn}, \binits{C.}}:
\batitle{Probing multimode squeezing with correlation functions}.
\bjtitle{New Journal of Physics}
\bvolume{13}(\bissue{3}),
\bfpage{033027}
(\byear{2011})
\end{barticle}
\endbibitem

%%% 46
\bibitem[\protect\citeauthoryear{Adcock et~al.}{2018}]{adcock2018hard}
\begin{barticle}
\bauthor{\bsnm{Adcock}, \binits{J.C.}},
\bauthor{\bsnm{Morley-Short}, \binits{S.}},
\bauthor{\bsnm{Silverstone}, \binits{J.W.}},
\bauthor{\bsnm{Thompson}, \binits{M.G.}}:
\batitle{Hard limits on the postselectability of optical graph states}.
\bjtitle{Quantum Science and Technology}
\bvolume{4}(\bissue{1}),
\bfpage{015010}
(\byear{2018})
\end{barticle}
\endbibitem

%%% 47
\bibitem[\protect\citeauthoryear{Werner}{1989}]{werner1989quantum}
\begin{barticle}
\bauthor{\bsnm{Werner}, \binits{R.F.}}:
\batitle{Quantum states with einstein-podolsky-rosen correlations admitting a hidden-variable model}.
\bjtitle{Physical Review A}
\bvolume{40}(\bissue{8}),
\bfpage{4277}
(\byear{1989})
\end{barticle}
\endbibitem

%%% 48
\bibitem[\protect\citeauthoryear{Sheng et~al.}{2009}]{sheng2009multipartite}
\begin{barticle}
\bauthor{\bsnm{Sheng}, \binits{Y.-B.}},
\bauthor{\bsnm{Deng}, \binits{F.-G.}},
\bauthor{\bsnm{Zhao}, \binits{B.-K.}},
\bauthor{\bsnm{Wang}, \binits{T.-J.}},
\bauthor{\bsnm{Zhou}, \binits{H.-Y.}}:
\batitle{Multipartite entanglement purification with quantum nondemolition detectors}.
\bjtitle{The European Physical Journal D}
\bvolume{55}(\bissue{1}),
\bfpage{235}--\blpage{242}
(\byear{2009})
\end{barticle}
\endbibitem

%%% 49
\bibitem[\protect\citeauthoryear{James et~al.}{2001}]{james2001measurement}
\begin{barticle}
\bauthor{\bsnm{James}, \binits{D.F.}},
\bauthor{\bsnm{Kwiat}, \binits{P.G.}},
\bauthor{\bsnm{Munro}, \binits{W.J.}},
\bauthor{\bsnm{White}, \binits{A.G.}}:
\batitle{Measurement of qubits}.
\bjtitle{Physical Review A}
\bvolume{64}(\bissue{5}),
\bfpage{052312}
(\byear{2001})
\end{barticle}
\endbibitem

\end{thebibliography}

\clearpage % 换页，开始第二篇

\newcommand{\papertitle}[2]{%
  \begin{center}
    {\LARGE #1\par} % 大写+加粗+居中
  \end{center}
  \vspace{1em}
}
% ===== 第二篇 =====
\papertitle{Supplementary Information: Chip-to-chip hyperentanglement distribution and entanglement purification using silicon integrated photonics}

% 1) 计数器重置
\setcounter{page}{1}
\setcounter{section}{0}
\setcounter{figure}{0}
\setcounter{table}{0}
\setcounter{equation}{0}

% 2) 打印出来加前缀：图表方程都从 S1 开始
\renewcommand{\thefigure}{S\arabic{figure}}
\renewcommand{\thetable}{S\arabic{table}}
\renewcommand{\theequation}{S\arabic{equation}}
% （若连节号也想带前缀，可用：\renewcommand{\thesection}{S\arabic{section}}）

% 3) 超链接锚点也加唯一前缀，避免跳到上一篇
\makeatletter
\renewcommand{\theHfigure}{paperB.S\arabic{figure}}
\renewcommand{\theHtable}{paperB.S\arabic{table}}
\renewcommand{\theHequation}{paperB.S\arabic{equation}}
\renewcommand{\theHsection}{paperB.sec.\arabic{section}}
\makeatother

\section{Chip and device details}
\begin{figure}[h]
		\mbox{%
			\includegraphics[width=13cm]{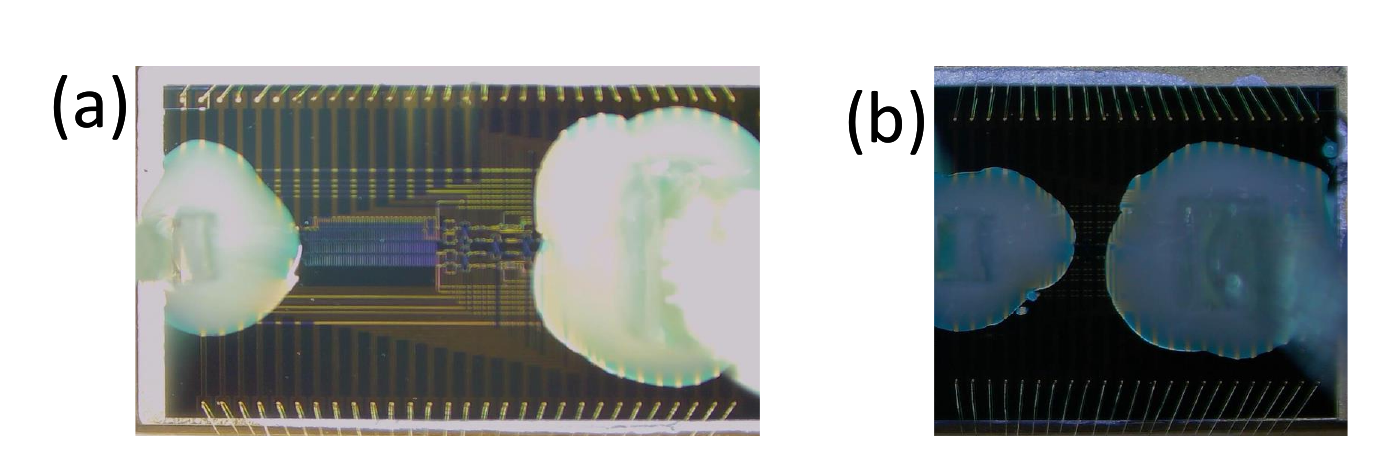}%
		}\caption{\label{supp1} The wire-bonded chip Charlie/Alice (a) and Bob (b). The electrical wires connect the pads on the chip to the pads on the PCB. The fiber array is aligned with the grating couplers on the chip and glued on chip, with all 2D GCs positioned on the left side and all 1D GCs on the right side.}
	\end{figure}

The chip is fabricated on a 250 nm silicon-on-insulator (SOI) platform using deep ultraviolet lithography (DUV). The wire-bonded chips are shown in Fig.~\ref{supp1}.

All devices on the chip operate in the TE mode. The waveguide loss is approximately 5 dB/cm, and each MMI introduces about 0.5 dB of loss. The coupling angles of the 1D GC and 2D GC are optimized to achieve the best coupling efficiency of -5.3 dB at 1550 nm for the 1D GC and -5.9 dB at 1560 nm for the 2D GC \cite{xue2019two}. The measured count rate of the signal photons on the time tagger is 500 kHz, while that of the idler photons is 3 kHz. Considering the coincidence rate of 10 Hz, the total detection loss is 24.8 dB for the signal photons and 47.0 dB for the idler photons (including detector efficiency).

%需不需要把图放上去
The free spectral range (FSR) of the AMZI we used in this work is measured to be 322.7~GHz, which means that the spectral position of the photons it separates is slightly offset from the designed signal (1539.77~nm) and idler (1558.98~nm) wavelengths. This mismatch results in around 1~dB of insertion loss for the signal photons.

We use on-chip titanium heaters with gold interconnects to connect the pads. The resistance of each titanium heater is measured to be $1000\Omega$, with a $V_{\pi}$ of approximately 2.5 V. The silicon dioxide at both sides of the titanium heaters is etched to form trenches, which reduces thermal crosstalk between heaters and improves thermal efficiency.

\section{Source information}
\begin{figure}[h]
		\mbox{%
			\includegraphics[width=13cm]{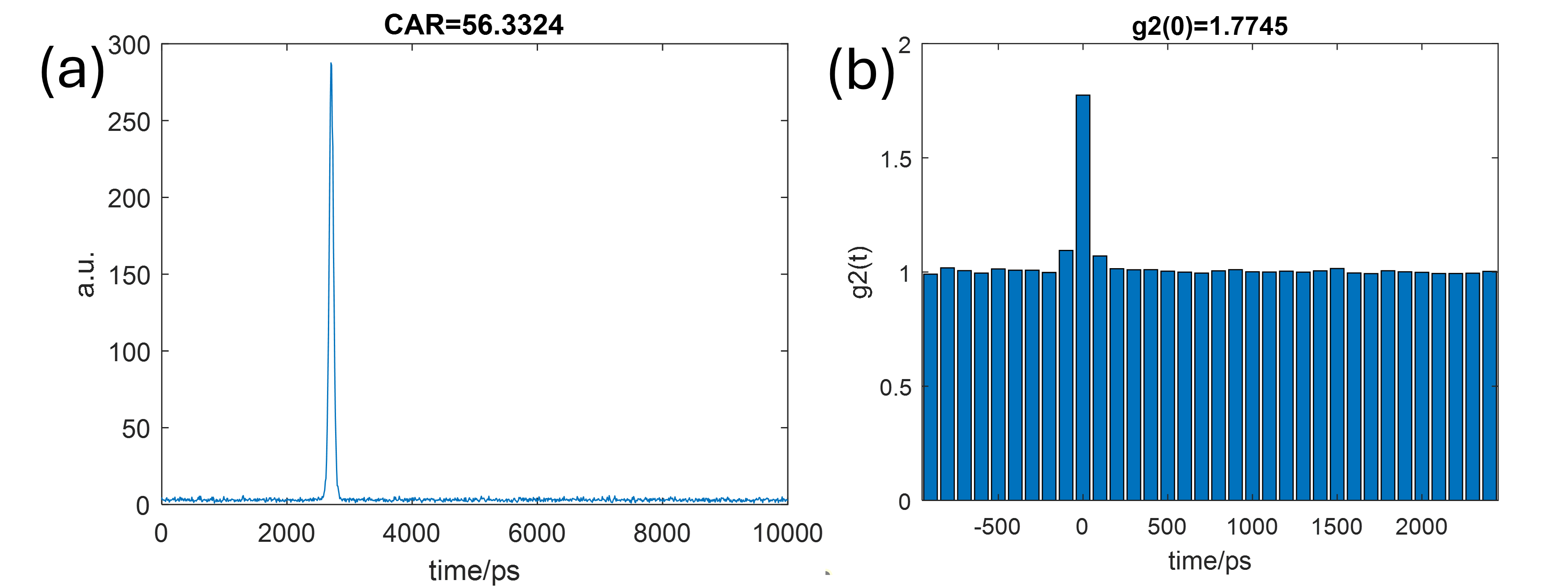}%
		}\caption{\label{supp2} The coincidence histogram for squeezing (a) and spectral purity (b) measurements.}
	\end{figure}

The feasibility of generating photon pairs from straight waveguides via SFWM has been demonstrated in previous works \cite{silverstone2014chip,wang2018multienta,adcock2019programmable}. Here, to evaluate the performance of the filtering system and ensure minimal multi-photon generation, we measured the squeezing parameter and the spectral fidelity, as shown in Fig.~\ref{supp2}.

In the measurement shown in Fig.~\ref{supp2}(a), all MZIs in the EGC, purification circuits, and measurement circuits are set to identity, and the coincidences between the signal and idler photons are directly recorded. From the coincidence histogram, we calculate the coincidence-to-accidental ratio (CAR) of the signal-idler. In our experiment, when the pump power before the chip is 18.9 dBm, we obtain a coincidence rate of 10.38 Hz and a CAR of 56.3. Using the CAR value, we calculate the squeezing parameter according to Eq.~\ref{eqsup1} \cite{danielthesis}.
\begin{equation}
\begin{alignedat}{1}
\frac{1}{CAR} &=\frac{cc(\Delta t\neq 0)}{cc(\Delta t=0)} = \frac{(1-a^2\xi^2)\xi^2}{(1+a\xi^2)(1-a\xi^2)}\\
a&=1-\eta
\end{alignedat}
\label{eqsup1}
\end{equation}
Here, as I mentioned in last section, the detection efficiency $\eta$ in our system is very low, so the parameter $a$ can be approximated as 1. The calculated squeezing parameter is $\xi = 0.02$, indicating that the multi-photon pair fraction is negligible compared with the single-photon pair generation.

In the measurement shown in Fig.~\ref{supp2}(b), an off-chip beam splitter (BS) is added to divide the signal photons into two paths, and the coincidence histogram is recorded to measure the unheralded $g^{(2)}(0)$ of the signal photons.
For a source based on SFWM or SPDC, the unheralded $g^{(2)}(0)$ value of the signal or idler photons is related to the Schmidt mode number $K$ of the joint spectral amplitude \cite{christ2011probing}, as shown in Eq. \ref{eqsup2}.
\begin{equation}
\begin{alignedat}{1}
g^{(2)}(0) &= 1 + \frac{1}{K}\\
P&\approx\frac{1}{K}
\end{alignedat}
\label{eqsup2}
\end{equation}
where $P$ is the spectral purity. This expression indicates that a higher $g^{(2)}(0)$ (maxmimum is 2 for SFWM) corresponds to a smaller number of spectral modes, implying higher spectral purity. 
For the $g^{(2)}(0)$ measurement shown in Fig.~\ref{supp2}(b), we obtain a raw spectral purity of 0.77. However, considering the dispersion of the coincidence histogram caused by the timing jitter of the SNSPDs and the time tagger (150 ps), we include the contributions from the two adjacent peaks. After this correction, the spectral purity is calculated to be 0.94.

\section{Purification protocol}
%1.完整的纠缠纯化推导，包括纠缠与片上纠缠的转化2.hardmard门3.从另外两个出口出来会怎么样4.一个表格，对于广泛的错误应该是怎么样的5.噪声模拟的时间分配6.理论分析下纠缠纯化最大能到多少，提高多少，画个图出来
\subsection{Conversion from on-chip path entanglement to fiber-based hyperentanglement}
One pair of 2D GCs can convert the on-chip path-encoded states $\ket{0}$ and $\ket{1}$ into fiber-based polarization-encoded states $\ket{H}$ and $\ket{V}$, respectively \cite{wang2018multienta}. With two pairs of fibers and 2D GCs, we can further introduce a fiber-based spatial-mode qubit. For either the signal or idler photon, the 2D GCs perform the mapping $\ket{0}\rightarrow\ket{0H}$, $\ket{1}\rightarrow\ket{0V}$, $\ket{2}\rightarrow\ket{1H}$, and $\ket{3}\rightarrow\ket{1V}$, where $\ket{0H}$ represents a horizontally polarized photon in the first fiber and $\ket{1V}$ represents a vertically polarized photon in the second fiber. After the four snake-shaped waveguides and AMZIs, the generated high-dimensional entangled state is converted as:
\begin{equation}
\begin{alignedat}{1}
\ket{\phi}=&\tfrac{1}{2}\left(\ket{00}+\ket{11}+\ket{22}+\ket{33} \right)\\
%&=\tfrac{1}{2}\left(\ket{0H}_s\ket{0H}_i+\ket{0V}_s\ket{0V}_i+\ket{1H}_s\ket{1H}_i+\ket{1V}_s\ket{1V}_i \right)\\
\xrightarrow{\text{2D GCs}}&\tfrac{1}{2}\left(\ket{0H0H}+\ket{0V0V}+\ket{1H1H}+\ket{1V1V} \right)\\
=&\tfrac{1}{2}\left(\ket{00}(\ket{HH}+\ket{VV}) +\ket{11}(\ket{HH}+\ket{VV})\right)\\
=&\tfrac{1}{2}\left(\ket{00}+\ket{11} \right)\left(\ket{HH}+\ket{VV} \right)=\ket{\phi^+} \otimes \ket{\Phi^+}
\end{alignedat}
\label{eqsup3}
\end{equation}
Here, $\ket{\phi^+} \otimes \ket{\Phi^+}$ represents the hyperentanglement in fiber. On chips Alice and Bob, the 2D GCs convert the hyperentanglement back into high-dimensional on-chip entanglement following the same mapping: $\ket{0H}\rightarrow\ket{0}$, $\ket{0V}\rightarrow\ket{1}$, $\ket{1H}\rightarrow\ket{2}$, and $\ket{1V}\rightarrow\ket{3}$.

\subsection{Entanglement purification based on integrated circuits}
In the purification circuit (orange dashed boxes) shown in Fig.~\ref{fig1}(b) and (c), the photons are swapped on chip for both the signal and idler photons according to the mapping $\ket{0}\rightarrow\ket{1}$, $\ket{1}\rightarrow\ket{3}$, $\ket{2}\rightarrow\ket{2}$, and $\ket{3}\rightarrow\ket{0}$. This operation equivalently transforms the fiber-based qubits as $\ket{0H}\rightarrow\ket{0V}$, $\ket{0V}\rightarrow\ket{1V}$, $\ket{1H}\rightarrow\ket{1H}$, and $\ket{1V}\rightarrow\ket{0H}$.

In the following discussion of this section, we use the hyperentangled qubit representation to illustrate how our circuit equivalently transforms the photon states. 
For the state without any error, our purification circuit transforms the state as:
\begin{equation}
\begin{alignedat}{1}
&\ket{\phi^+} \otimes \ket{\Phi^+}=\tfrac{1}{2}\left(\ket{00}+\ket{11} \right)\left(\ket{HH}+\ket{VV} \right)\\
=&\tfrac{1}{2}\left(\ket{0H0H}+\ket{0V0V}+\ket{1H1H}+\ket{1V1V} \right)\\
\xrightarrow{\text{Puri}}&\tfrac{1}{2}\left(\ket{0V0V}+\ket{1V1V}+\ket{1H1H}+ \ket{0H0H}\right)\\
=&\tfrac{1}{2}\left(\ket{00}+\ket{11}+\ket{22}+\ket{33} \right)
\end{alignedat}
\label{eqsup4}
\end{equation}
From Eq.~\ref{eqsup4}, we can see that after purification, the sequence of the four terms is simply swapped, and the resulting state remains the original entangled state. In detection, we only collect photons from the first two waveguides (in red circle) for both the signal and idler photons, so the final remaining state is $\tfrac{1}{\sqrt{2}}\left(\ket{00}+\ket{11}\right)$.

For the state without one BF error on polarization or spatial-mode qubit, our purification circuit transforms the state as:
\begin{equation}
\begin{alignedat}{1}
&\ket{\psi^+} \otimes \ket{\Phi^+}=\tfrac{1}{2}\left(\ket{01}+\ket{10} \right)\left(\ket{HH}+\ket{VV} \right)\\
=&\tfrac{1}{2}\left(\ket{0H1H}+\ket{0V1V}+\ket{1H0H}+\ket{1V0V} \right)\\
\xrightarrow{\text{Puri}}&\tfrac{1}{2}\left(\ket{0V1H}+\ket{1V0H}+\ket{1H0V}+ \ket{0H1V}\right)\\
=&\tfrac{1}{2}\left(\ket{12}+\ket{30}+\ket{21}+\ket{03} \right)
\end{alignedat}
\label{eqsup5}
\end{equation}

\begin{equation}
\begin{alignedat}{1}
&\ket{\phi^+} \otimes \ket{\Psi^+}=\tfrac{1}{2}\left(\ket{00}+\ket{11} \right)\left(\ket{HV}+\ket{VH} \right)\\
=&\tfrac{1}{2}\left(\ket{0H0V}+\ket{0V0H}+\ket{1H1V}+\ket{1V1H} \right)\\
\xrightarrow{\text{Puri}}&\tfrac{1}{2}\left(\ket{0V1V}+\ket{1V0V}+\ket{1H0H}+ \ket{0H1H}\right)\\
=&\tfrac{1}{2}\left(\ket{13}+\ket{31}+\ket{20}+\ket{02} \right)
\end{alignedat}
\label{eqsup6}
\end{equation}

Since we only collect photons from the first two waveguides, all four terms in the results of Eq.~\ref{eqsup5} and Eq.~\ref{eqsup6} contain only one detected photon. More explicitly, they can be written as $\tfrac{1}{2}\left(\ket{1,\text{None}}+\ket{\text{None},0}+\ket{\text{None},1}+\ket{0,\text{None}}\right)$ and $\tfrac{1}{2}\left(\ket{1,\text{None}}+\ket{\text{None},1}+\ket{\text{None},0}+\ket{0,\text{None}}\right)$. In entanglement experiments, coincidences are recorded, so states with a single BF error in either the polarization or spatial-mode qubit do not affect the measured data. It is important to note that this selection of output waveguides differs from the post-selection typically used in constructing path-encoded CNOT gates \cite{adcock2018hard}, because in this work, we do not select data based on specific detection events and all coincident detections correspond to the purified photons.

For photon pairs with two BF errors on both the polarization and spatial-mode qubits, our purification circuit transforms the state as:
\begin{equation}
\begin{alignedat}{1}
&\ket{\psi^+} \otimes \ket{\Psi^+}=\tfrac{1}{2}\left(\ket{01}+\ket{10} \right)\left(\ket{HV}+\ket{VH} \right)\\
=&\tfrac{1}{2}\left(\ket{0H1V}+\ket{0V1H}+\ket{1H0V}+\ket{1V0H} \right)\\
\xrightarrow{\text{Puri}}&\tfrac{1}{2}\left(\ket{0V0H}+\ket{1H1V}+\ket{1H1V}+ \ket{1V1H}\right)\\
=&\tfrac{1}{2}\left(\ket{01}+\ket{10}+\ket{23}+\ket{32} \right)
\end{alignedat}
\label{eqsup7}
\end{equation}
From Eq.~\ref{eqsup7}, we can see that for the state with BF errors on both the polarization and spatial-mode qubits, the photons collected from the first two waveguides correspond to the state $\tfrac{1}{\sqrt{2}}\left(\ket{01}+\ket{10}\right)$. Although coincidence still occurs, the resulting state is not the desired one. This indicates that our purification circuit cannot correct states suffering from simultaneous BF errors on both the polarization and spatial-mode qubits. Fortunately, the probability of BF errors occurring on both qubits is much lower than that on a single qubit \cite{puri32021}, as we will discuss in the next section.

Here, if we collect both the signal and idler from the third and fourth waveguides, the purified outcomes map to error syndromes as follows:
\begin{equation}
\begin{alignedat}{1}
&\ket{\phi^+} \otimes \ket{\Phi^+}
\xrightarrow{\text{Puri}}\tfrac{1}{\sqrt{2}}\left(\ket{22}+\ket{33}\right)\\
&\ket{\psi^+} \otimes \ket{\Phi^+}
\xrightarrow{\text{Puri}}\tfrac{1}{2}\left(\ket{3,\text{None}}+\ket{\text{None},2}+\ket{\text{None},3}+\ket{2,\text{None}}\right)\\
&\ket{\phi^+} \otimes \ket{\Psi^+}
\xrightarrow{\text{Puri}}\tfrac{1}{2}\left(\ket{3,\text{None}}+\ket{\text{None},3}+\ket{\text{None},2}+\ket{2,\text{None}}\right)\\
&\ket{\psi^+} \otimes \ket{\Psi^+}
\xrightarrow{\text{Puri}}\tfrac{1}{\sqrt{2}}\left(\ket{23}+\ket{32}\right)
\end{alignedat}
\label{eqsup8}
\end{equation}
We can see that the states with a single bit-flip error on either the polarization or spatial-mode qubit are removed, and the remaining state is purified. In fact, these photons can be collected in parallel with those from the first and second waveguides, yielding two groups of purified photons without spatial-mode encoding. This indicates that the spatial-mode DOF is consumed to purify the photons, or equivalently, the spatial mode is used to protect the polarization qubit.

\subsection{BF error to PF error conversion}
Since our purification circuit only swaps photons without changing their phases, it cannot purify photons with PF errors. For fiber-based hyperentangled qubits, Hadamard gates need to be applied to both degrees of freedom of the two photons to convert the PF error to BF error.
The Hadamard gate maps $\ket{0}$ to $(\ket{0}+\ket{1})/\sqrt{2}$ and $\ket{1}$ to $(\ket{0}-\ket{1})/\sqrt{2}$.
Taking $\ket{\phi^+} \otimes \ket{\Phi^-}$ as an example:

\begin{equation}
\begin{alignedat}{1}
&\ket{\phi^+} \otimes \ket{\Phi^-}=\tfrac{1}{2}\left(\ket{00}+\ket{11} \right)\left(\ket{HH}-\ket{VV} \right)\xrightarrow{\text{Hadamard}}\\
&\tfrac{1}{2}\left(\frac{\scriptstyle(\ket{0}+\ket{1})(\ket{0}+\ket{1})+(\ket{0}-\ket{1})(\ket{0}-\ket{1})}{\scriptstyle2}\right)\left(\frac{\scriptstyle(\ket{H}+\ket{V})(\ket{H}+\ket{V})-(\ket{H}-\ket{V})(\ket{H}-\ket{V})}{\scriptstyle2}\right)\\
=&\tfrac{1}{2}\left(\ket{00}+\ket{11} \right)\left(\ket{HV}+\ket{VH} \right)\\
=&\ket{\phi^+}\otimes\ket{\Psi^+}\, .
\end{alignedat}
\label{eqsup9}
\end{equation}

In this work, as shown in the circuit in Fig.~\ref{fig3}(k), we first apply Hadamard gates on waveguide modes $\ket{0}$ and $\ket{1}$ for both the signal and idler, which transform as $\ket{0}\rightarrow(\ket{0}+\ket{1})/\sqrt{2}$ and $\ket{1}\rightarrow(\ket{0}-\ket{1})/\sqrt{2}$. Similarly, Hadamard gates are applied on $\ket{2}$ and $\ket{3}$, giving $\ket{2}\rightarrow(\ket{2}+\ket{3})/\sqrt{2}$ and $\ket{3}\rightarrow(\ket{2}-\ket{3})/\sqrt{2}$. Then, using the waveguide crossings, we apply Hadamard gates on $\ket{0}$ and $\ket{2}$, and on $\ket{1}$ and $\ket{3}$, which transform as $\ket{0}\rightarrow(\ket{0}+\ket{2})/\sqrt{2}$, $\ket{2}\rightarrow(\ket{0}-\ket{2})/\sqrt{2}$, $\ket{1}\rightarrow(\ket{1}+\ket{3})/\sqrt{2}$, and $\ket{3}\rightarrow(\ket{1}-\ket{3})/\sqrt{2}$, as shown below:

\begin{equation}
\begin{alignedat}{1}
&\ket{\phi^+} \otimes \ket{\Phi^-}=\tfrac{1}{2}\left(\ket{00}+\ket{11} \right)\left(\ket{HH}-\ket{VV} \right)\\
=&\tfrac{1}{2}\left(\ket{0H0H}-\ket{0V0V}+\ket{1H1H}-\ket{1V1V} \right)\xrightarrow{\text{2D GC}}\\
&\tfrac{1}{2}\left(\ket{00}-\ket{11}+\ket{22}-\ket{33} \right)\xrightarrow{\text{1st level Hadamard}}\\
=&\tfrac{1}{2}\left(\frac{\scriptstyle(\ket{0}+\ket{1})(\ket{0}+\ket{1})-(\ket{0}-\ket{1})(\ket{0}-\ket{1})}{\scriptstyle2}+ \frac{\scriptstyle(\ket{2}+\ket{3})(\ket{2}+\ket{3})-(\ket{2}-\ket{3})(\ket{2}-\ket{3})}{\scriptstyle2}\right)\\
=&\tfrac{1}{2}\left(\ket{01}+\ket{10}+\ket{23}+\ket{32} \right)\xrightarrow{\text{2nd level Hadamard}}\\
&\tfrac{1}{2}\left(\frac{\scriptstyle(\ket{0}+\ket{2})(\ket{1}+\ket{3})+(\ket{1}+\ket{3})(\ket{0}+\ket{2})}{\scriptstyle2}+ \frac{\scriptstyle(\ket{0}-\ket{2})(\ket{1}-\ket{3})+(\ket{1}-\ket{3})(\ket{0}-\ket{2})}{\scriptstyle2}\right)\\
=&\tfrac{1}{2}\left(\ket{01}+\ket{10}+\ket{23}+\ket{32}\right) \\
=&\tfrac{1}{2}\left(\ket{00}+\ket{11} \right)\left(\ket{HV}+\ket{VH} \right)\\
=&\ket{\phi^+}\otimes\ket{\Psi^+}\, .
\end{alignedat}
\label{eqsup10}
\end{equation}

The example used here is a PF on the polarization bit, so the first layer of Hadamard gates converts it into a BF. The second layer of Hadamard gates implemented with waveguide crossings should not affect the state, as shown in Eq.~\ref{eqsup10}.

In this way, with the PF EGC, we can simulate the PF error, and using the Hadamard gates and purification circuits, we can purify the PF error.

\subsection{Time distribution of no error, single-bit error and two-bit errors}
\begin{figure}[h]
    \centering
    \includegraphics[width=12cm]{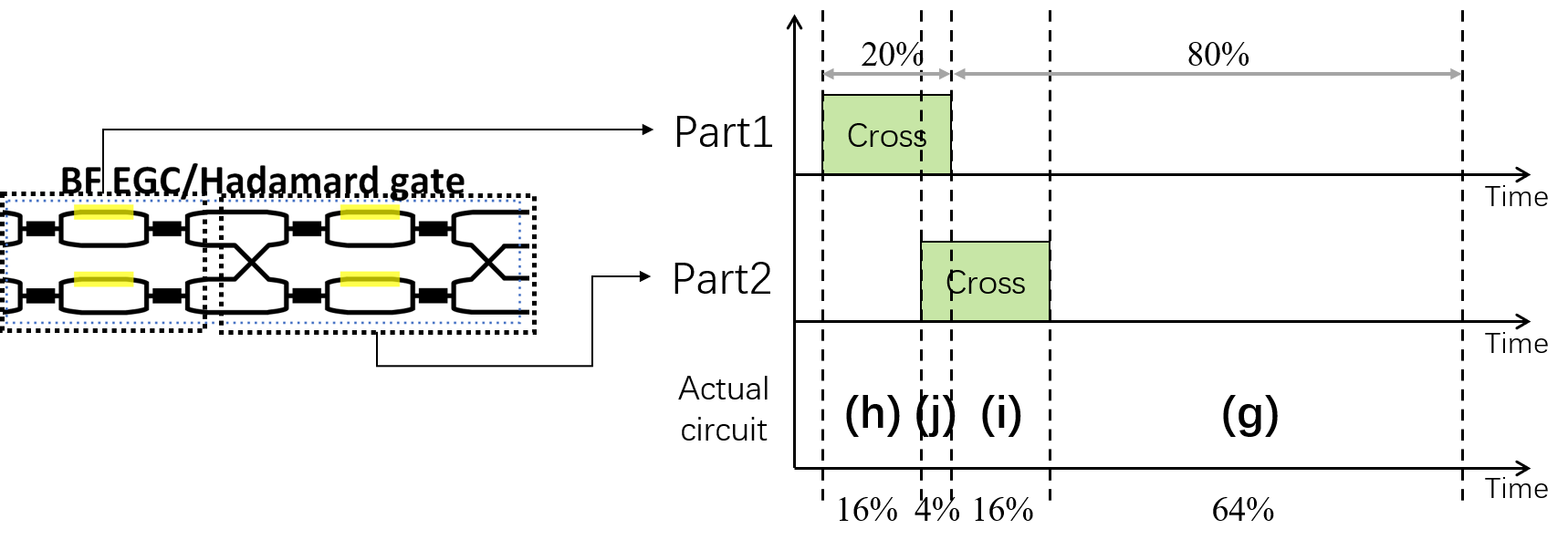}
    \caption{The method we simulate the 20\% BF error in a time distribution is same as Hu's work \cite{puri32021}. }
    \label{timedistrerror}
\end{figure}

As we mentioned in the text, to simulate the BF error in noisy fiber channels, a BF EGC at the Bob's part is designed and reconfigured to the circuits as shown in Fig.~\ref{fig3}(g)-\ref{fig3}(j) in a time distribution to satisfy a 20\% BF rate in the polarization and spatial dimension \cite{puri32021}, as shown in Fig. \ref{timedistrerror}. 

In a fiber channel, a 20\% BF error means that both the spatial and polarization bits independently experience a 20\% probability of flipping. As shown in Fig.~\ref{timedistrerror}, in the experiment, I configured the first two MZIs to switch to the cross state 20\% of the time, and similarly, the second pair of MZIs also switched to the cross state 20\% of the time. We set the timing positions of the two types of errors to create a 4\% overlap, so that both errors occur simultaneously with a 4\% probability.
As a result, 64\% of the time there is no error, 16\% of the time there is the BF error on either the spatial or polarization bit, and 4\% of the time there is both BF errors.

It is important to note that to correctly simulate the purification process, the MZI should be configured to alternate between the flip and non-flip states over time, rather than being fixed at a 20\% splitting ratio throughout the experiment. 
This is because purification aims to purify a mixed state. If the MZI changes the state to another pure state, it can always be corrected by applying a unitary.

\subsection{Theoretical prediction of purification result}
If the fidelity after a noisy channel for the polarization qubit is denoted as $F_1$, and that for the spatial-mode qubit is $F_2$, when there is only BF error, the fidelity after purification $F'$ is given by~\cite{puri32021}

\begin{equation}
\begin{alignedat}{1}
F'=\frac{F_1F_2}{F_1F_2+(1-F_1)(1-F_2)}
\end{alignedat}
\label{eqsup11}
\end{equation}

\begin{figure}[h]
    \centering
    \includegraphics[width=6cm]{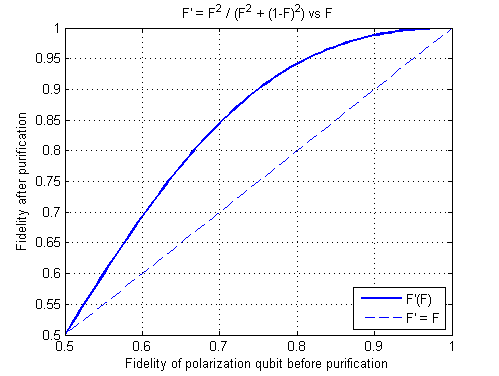}
    \caption{\label{supp4} A theoretical simulation of the fidelity before and after purification when there is only BF error.}
\end{figure}

Here we consider a symmetric channel with $F_1 = F_2$. The relationship between the fidelity after purification $F'$ and the fidelity before purification $F$ is shown in Fig.~\ref{supp4}.

From Fig.~\ref{supp4} and Eq.~\ref{eqsup11}, we can see that as long as $F > 0.5$, the purified fidelity $F'$ is greater than $F$. When the initial fidelity is $F = 0.75$, the purification achieves the maximum improvement, with the purified state reaching a fidelity of $F'=0.9$, corresponding to an improvement of 0.15. In our case, we set the BF rate to 0.2, representing an initial fidelity of $F = 0.8$, and the improvement after purification is 0.14.

\subsection{Entanglement purification against white noise}
In practice, we cannot assume that the channel is dominated by BF noise. In fact, both BF and PF noise can occur independently. 
As Werner's theory \cite{werner1989quantum}, a imperfect entangled states are described as a mixed state using the following density matrix:
\begin{equation}
\begin{alignedat}{1}
\rho = F\ket{\Phi^+}\bra{\Phi^+} 
+ \frac{1-F}{3}\ket{\Phi^-}\bra{\Phi^-}
+ \frac{1-F}{3}\ket{\Psi^+}\bra{\Psi^+}
+ \frac{1-F}{3}\ket{\Psi^-}\bra{\Psi^-}
\end{alignedat}
\label{eqsup12}
\end{equation}

The state described in Eq.~\ref{eqsup12} is known as a Werner state, which can be regarded as a mixture of $\ket{\Phi^+}$ and a maximally mixed state (white noise). Here, we present a table to illustrate how the purification process works in a channel with white noise, as shown in Table~\ref{table1}.

%1.可能要换wener state2.postselection要讲一讲，也讲下CNOT，就是cnot加postselect3.上面那个不是时变的，而是纯态混态

\begin{table}[h]
\centering
\caption{\label{table1}The table lists the probabilities of BF and PF errors occurring and whether the errored states still cause coincidences after purification. Spatial (Before) and Polar (Before) denote the spatial-mode bit and polarization bit in the fiber and their probability, respectively. HD state (After) represents the path-encoded high-dimensional state on the chip after the 2D GCs and purification circuit.}
\begin{tabular}[htbp]{@{}lllll@{}}
\hline
Spatial (Before) & Polar (Before) & HD state (After) & Probability & Coincidence? \\
\hline
$\ket{\phi^+}$\ $F$ & $\ket{\Phi^+}$\ $F$ & $\ket{11}+\ket{33}+\ket{22}+\ket{00}$ & $F^2$ & Yes \\
$\ket{\phi^+}$\ $F$ & $\ket{\Phi^-}$\ $\frac{1-F}{3}$ & $\ket{11}-\ket{33}+\ket{22}-\ket{00}$ & $\frac{F(1-F)}{3}$ & Yes \\
$\ket{\phi^+}$\ $F$ & $\ket{\Psi^+}$\ $\frac{1-F}{3}$ & $\ket{13}+\ket{31}+\ket{20}+\ket{02}$ & $\frac{F(1-F)}{3}$ & No \\
$\ket{\phi^+}$\ $F$ & $\ket{\Psi^-}$\ $\frac{1-F}{3}$ & $\ket{13}-\ket{31}+\ket{20}-\ket{02}$ & $\frac{F(1-F)}{3}$ & No \\[2pt]
$\ket{\phi^-}$\ $\frac{1-F}{3}$ & $\ket{\Phi^+}$\ $F$ & $\ket{11}+\ket{33}-\ket{22}-\ket{00}$ & $\frac{F(1-F)}{3}$ & Yes \\
$\ket{\phi^-}$\ $\frac{1-F}{3}$ & $\ket{\Phi^-}$\ $\frac{1-F}{3}$ & $\ket{11}-\ket{33}-\ket{22}+\ket{00}$ & $\left(\frac{1-F}{3}\right)^2$ & Yes \\
$\ket{\phi^-}$\ $\frac{1-F}{3}$ & $\ket{\Psi^+}$\ $\frac{1-F}{3}$ & $\ket{13}+\ket{31}-\ket{20}-\ket{02}$ & $\left(\frac{1-F}{3}\right)^2$ & No \\
$\ket{\phi^-}$\ $\frac{1-F}{3}$ & $\ket{\Psi^-}$\ $\frac{1-F}{3}$ & $\ket{13}-\ket{31}-\ket{20}+\ket{02}$ & $\left(\frac{1-F}{3}\right)^2$ & No \\[2pt]
$\ket{\psi^+}$\ $\frac{1-F}{3}$ & $\ket{\Phi^+}$\ $F$ & $\ket{12}+\ket{30}+\ket{21}+\ket{03}$ & $\frac{F(1-F)}{3}$ & No \\
$\ket{\psi^+}$\ $\frac{1-F}{3}$ & $\ket{\Phi^-}$\ $\frac{1-F}{3}$ & $\ket{12}-\ket{30}+\ket{21}-\ket{03}$ & $\left(\frac{1-F}{3}\right)^2$ & No \\
$\ket{\psi^+}$\ $\frac{1-F}{3}$ & $\ket{\Psi^+}$\ $\frac{1-F}{3}$ & $\ket{10}+\ket{32}+\ket{23}+\ket{01}$ & $\left(\frac{1-F}{3}\right)^2$ & Yes \\
$\ket{\psi^+}$\ $\frac{1-F}{3}$ & $\ket{\Psi^-}$\ $\frac{1-F}{3}$ & $\ket{10}-\ket{32}+\ket{23}-\ket{01}$ & $\left(\frac{1-F}{3}\right)^2$ & Yes \\[2pt]
$\ket{\psi^-}$\ $\frac{1-F}{3}$ & $\ket{\Phi^+}$\ $F$ & $\ket{12}+\ket{30}-\ket{21}-\ket{03}$ & $\frac{F(1-F)}{3}$ & No \\
$\ket{\psi^-}$\ $\frac{1-F}{3}$ & $\ket{\Phi^-}$\ $\frac{1-F}{3}$ & $\ket{12}-\ket{30}-\ket{21}+\ket{03}$ & $\left(\frac{1-F}{3}\right)^2$ & No \\
$\ket{\psi^-}$\ $\frac{1-F}{3}$ & $\ket{\Psi^+}$\ $\frac{1-F}{3}$ & $\ket{10}+\ket{32}-\ket{23}-\ket{01}$ & $\left(\frac{1-F}{3}\right)^2$ & Yes \\
$\ket{\psi^-}$\ $\frac{1-F}{3}$ & $\ket{\Psi^-}$\ $\frac{1-F}{3}$ & $\ket{10}-\ket{32}-\ket{23}+\ket{01}$ & $\left(\frac{1-F}{3}\right)^2$ & Yes \\

\hline
\end{tabular}
\end{table}

In the simulation of Table~\ref{table1}, we still collect photons only from the first two waveguides, $\ket{0}$ and $\ket{1}$. Therefore, states such as $\ket{13}$, $\ket{30}$, $\ket{02}$, and $\ket{21}$ do not produce coincidences. Only photon pairs that cause coincidences are included in the dataset and used in the fidelity calculation.
Only the state $\ket{00}+\ket{11}$ is the desired one, so we calculate the fidelity of the final mixed state $F'$ with respect to $\ket{00}+\ket{11}$:

\begin{equation}
\begin{alignedat}{1}
F'=\frac{F^2+\tfrac{1}{9}(1-F)^2}{F^2+\tfrac{2}{3}F(1-F)+\tfrac{5}{9}(1-F)^2}
\end{alignedat}
\label{eqsup13}
\end{equation}

Eq.~\ref{eqsup13} corresponds to the result of the first purification protocol proposed by Bennett \cite{puri41996}. The purification based on hyperentanglement \cite{puri32021} is essentially a one-step implementation of Bennett's scheme, where different DOFs are used to purify the photons. This approach eliminates the need for two-photon interference and thus achieves very high entanglement efficiency. However, its limitation is that the purified photons cannot be further purified within the same circuit unless quantum nondemolition measurement is available \cite{sheng2009multipartite}.

\section{Phase locking system}
In hyperentanglement distribution, it is essential to lock the relative phase between the two spatial qubits. The technique used to achieve this is called an OPLL, which was originally developed for coherent optical communication, where the local oscillator laser and the signal laser must remain phase-locked after long-distance transmission.

An OPLL consists of three main components: a phase detector, a loop filter, and an optical voltage-controlled oscillator. The phase detector measures the phase difference between the signal and the local oscillator lasers. This phase error is passed to the loop filter, which adjusts the voltage applied to the phase shifter to compensate for the error. A basic control rule is to apply a larger correction when a larger phase difference is detected. If the direction of correction is properly set, the PLL will lock the phase of the signal laser to that of the local oscillator with small fluctuations. For optimal performance, a proportional-integral-derivative (PID) controller can be employed.

\begin{figure*}[h]
    \centering
    \includegraphics[width=13cm]{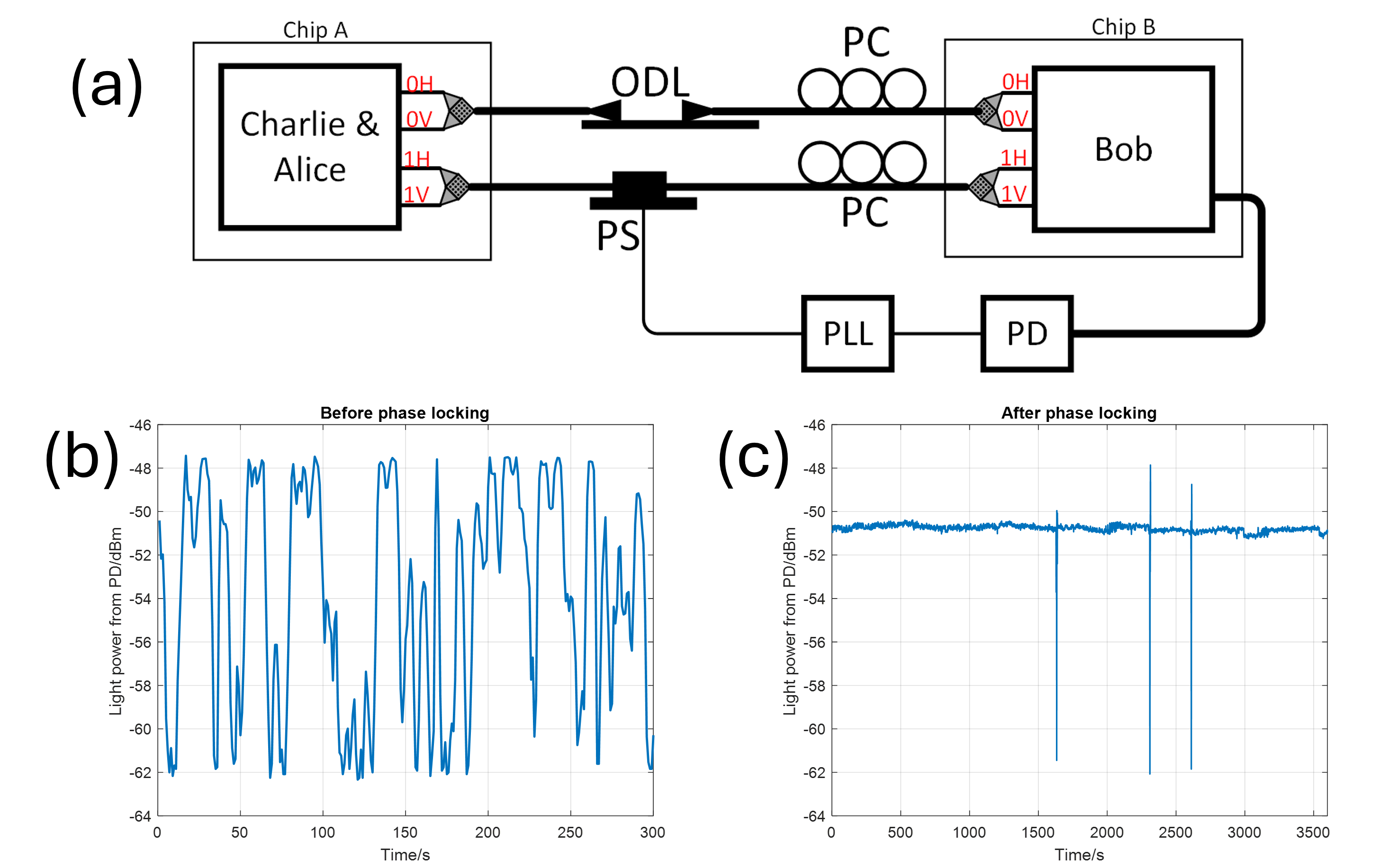}
    \caption{The optical PLL system and results in this work. ODL: optical tunable delay line; PS: electrical phase shifter; PLL: phase-lock control board; PD: photodetector; PC: polarization controller. (a) Experimental setup of the OPLL. (b) Power received on the PD over 5 minutes before phase locking is applied, showing random fluctuations. (c) Power received on the PD over 1 hour after phase locking is applied, showing stable interference.}
    \label{opll}
\end{figure*}

In our setup, as shown in Fig.~\ref{opll}(a), the on-chip MMI and the off-chip PD together serve as the phase detector. The PLL control board functions as the loop filter. It reads the power signal from the PD and generates a control voltage to drive the phase shifter (PS), which acts as an optical voltage-controlled oscillator. The response of the PLL can be tuned using PID parameters set on the computer. A tunable free space optical delay line ensures that the two fibers have precisely matched time delays to achieve optimal interference. Two PCs are used to ensure that the light coupled into Chip B has the correct polarization state.

From Fig.~\ref{opll}(b), we observe an interference contrast higher than 15~dB, indicating good temporal overlap between the two optical pulses in two fibers. The Fig.~\ref{opll}(b) also shows that the phase fluctuates on a timescale of about 10 seconds without phase locking. The locking point is set at half of the maximum interference power, where the PLL exhibits optimal performance.
After enabling the phase lock, as shown in Fig.~\ref{opll}(c), the power remains stable at $-51.2 \pm 0.2$~dBm, indicating that the relative phase between the two fibers is successfully locked with a relative power fluctuation of $\pm4.6\%$. Occasional phase unlocking events are observed, but the system typically get back to locks within 2 seconds. In the experiment, since the power is continuously monitored on a computer, data collection is paused during such brief unlock events to avoid affecting the final results.

\section{Quantum state tomography and CHSH measurement}
\begin{figure}[h!]
    \centering
    \includegraphics[width=13.1cm]{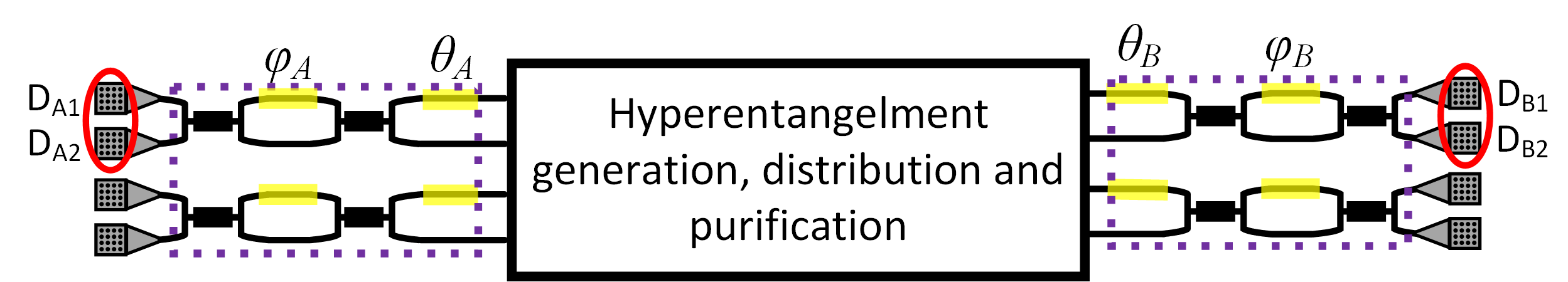}
    \caption{\label{qstmzi} The on-chip circuit for QST and CHSH inequality measurement.}
\end{figure}

In this work, we tune the phase of the MZIs at Alice and Bob to perform two-qubit QST and measure the CHSH inequality, as shown in Fig.~\ref{qstmzi}.

\subsection{Quantum state tomography}
\begin{figure}[h!]
    \centering
    \includegraphics[width=12.9cm]{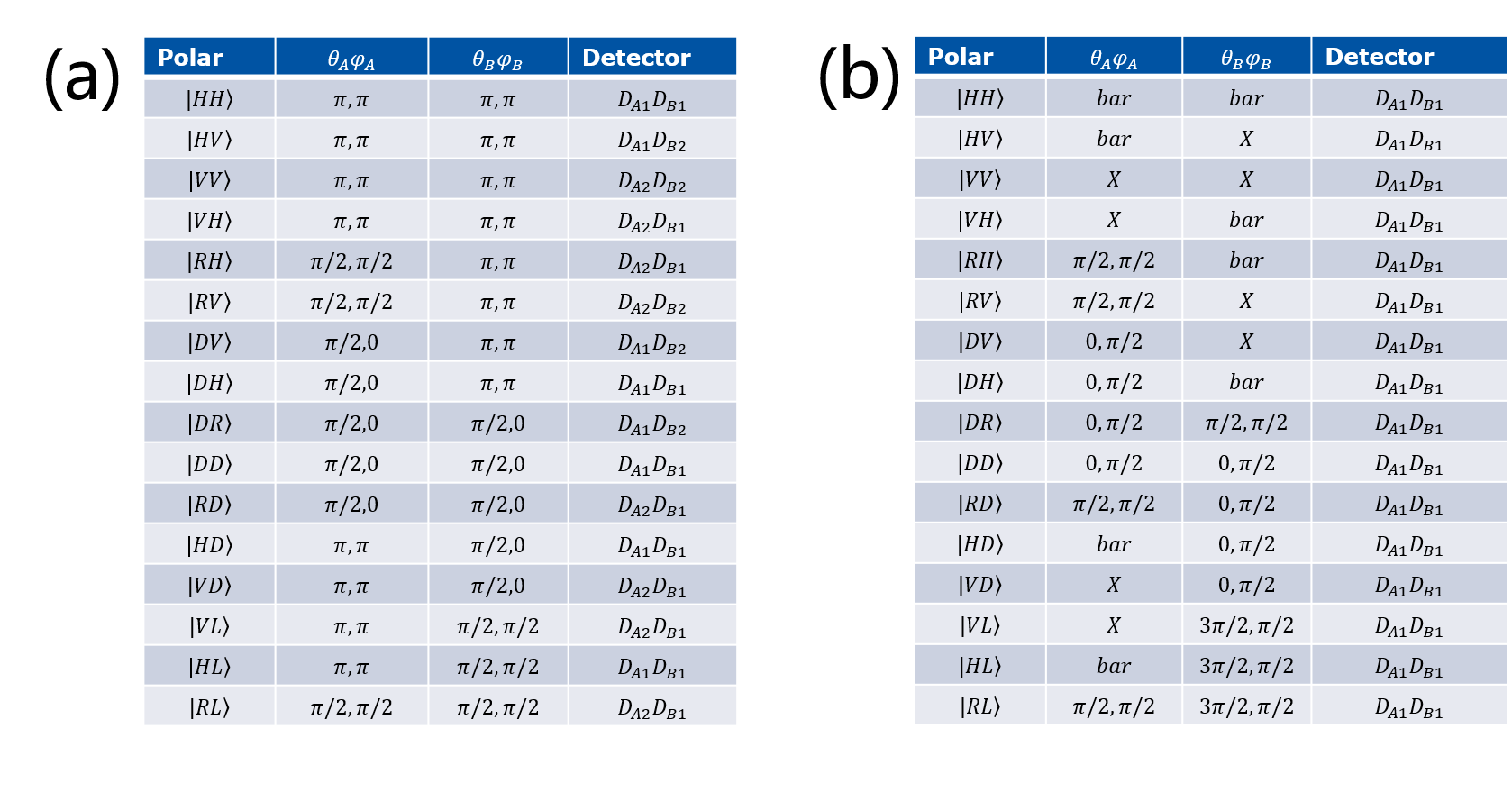}
    \caption{\label{qstbasis} The phase lookup table for QST. (a) Using four detectors. (b) Using two detectors.}
\end{figure}

The two-qubit QST \cite{james2001measurement} is used to reconstruct the density matrix of the quantum state and calculate its fidelity. By tuning the phase values $\theta_A$, $\psi_A$, $\theta_B$, and $\psi_B$ of the two MZIs shown in Fig.~\ref{qstmzi}, we can construct sixteen measurement bases.
The sixteen recorded two-fold coincidences from the four detectors are used to reconstruct the density matrix.
The corresponding sixteen phase combinations are shown in Fig.~\ref{qstbasis}.

If we denote the 16 coincidence counts obtained from Fig.~\ref{qstbasis}(a) as $n_{\nu}$ with $\nu = 1, 2, \ldots, 16$, the density matrix $\hat{\rho}$ can be calculated using the formula \cite{james2001measurement}:
\begin{equation}
\hat{\rho} = \left( \sum_{\nu=1}^{16} \hat{M}_{\nu} n_{\nu} \right) \bigg/ \left( \sum_{\nu=1}^{4} n_{\nu} \right),
\label{neweq101}
\end{equation}
where $\hat{M}_{\nu}$ is a set of matrices listed in James' original work \cite{james2001measurement}.

Once the density matrix $\hat{\rho}$ is reconstructed, the fidelity with respect to the target density matrix $\hat{\rho_0}$ can be calculated as:
\begin{equation}
F =\left(Tr\sqrt{\sqrt{\hat{\rho_0}}\hat{\rho}\sqrt{\hat{\rho_0}}} \right)^2.
\label{neweq102}
\end{equation}

In Fig.~\ref{qstbasis}(b), we also list the required phase settings for the 16 measurement bases assuming only two detectors are used. Here, "bar" indicates that the corresponding MZI phase is set to $\psi = \pi$, and "X" indicates $\psi = 0$. The phase parameter $\theta$ is irrelevant when the MZI is configured in the bar or cross state, so it can be omitted in those cases.

\subsection{Clauser-Horne-Shimony-Holt (CHSH) inequality}
\begin{figure}[h]
    \centering
    \includegraphics[width=12.9cm]{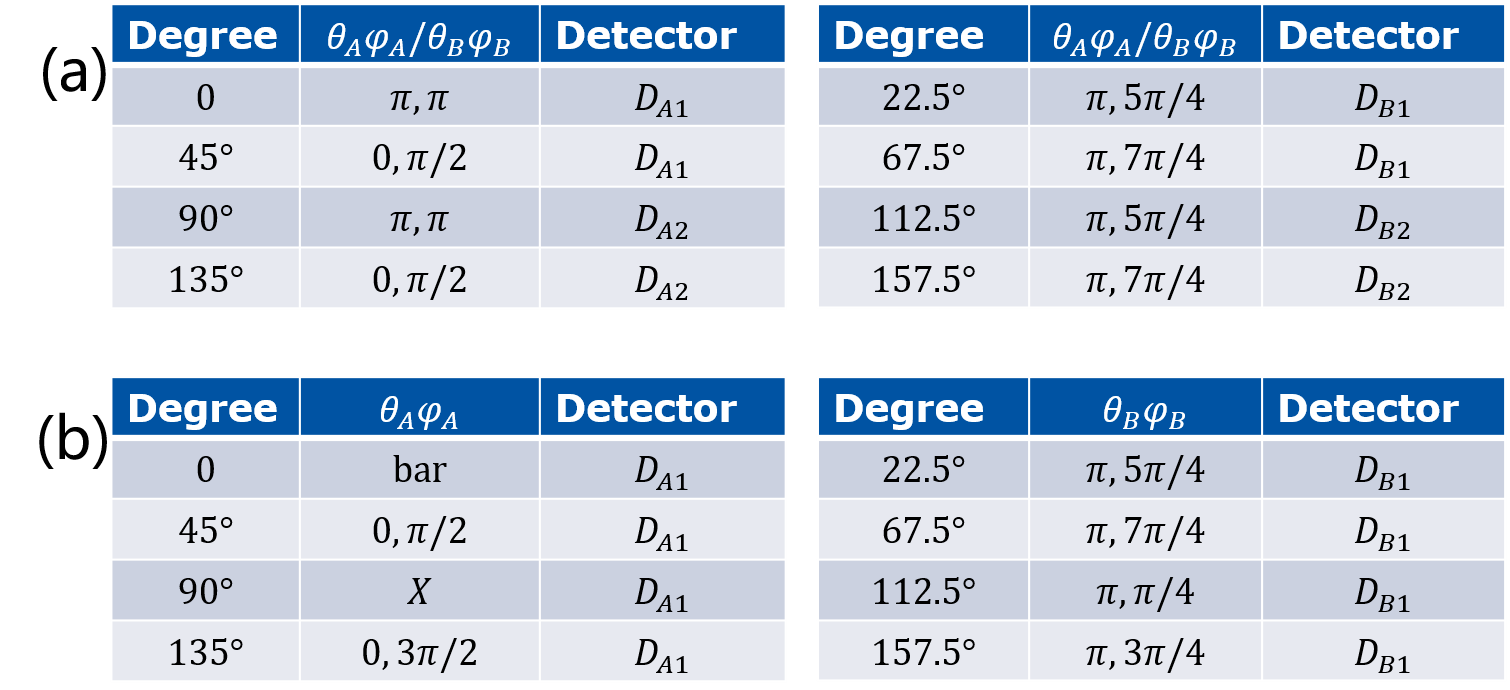}
    \caption{\label{chshbasis} The phase lookup table for CHSH inequality measurement. (a) Using four detectors. (b) Using two detectors.}  
\end{figure}
Breaking the CHSH inequality provides evidence that quantum nonlocality truly exists \cite{CHSH1}. In the measurement, suppose we have a polarization-entangled photon source. To measure the CHSH inequality, the two photons are sent separately to Alice and Bob, who detect the photons using polarizers placed in front of their detectors. We perform coincidence measurements of the entangled photon pairs when the polarizers are set at different angles. The CHSH inequality is given by:
\begin{equation}
|S| = |E(a, b) - E(a, b') + E(a', b) + E(a', b')| \leq 2
\label{neweq104}
\end{equation}
Here, $a$ and $a'$ are two setting angles of Alice's polarizer, and $b$ and $b'$ are those of Bob. $E(a,b)$ is the correlation expectation value under the settings $a$ and $b$, defined as:
\begin{equation}
E(a, b) = \frac{cc(a,b) -cc(a,b_\perp) - cc(a_\perp,b) + cc(a_\perp,b_\perp)}{cc(a,b) +cc(a,b_\perp) + cc(a_\perp,b) + cc(a_\perp,b_\perp)}
\end{equation}
where $cc(a,b)$ is the coincidence count when Alice's polarizer is set at angle $a$ and Bob's polarizer is set at angle $b$. The symbol $a_\perp$ denotes the angle perpendicular to $a$. 
In this work, we set the measurement angles as $a = 0^\circ$, $a' = 45^\circ$, $b = 22.5^\circ$, and $b' = 67.5^\circ$. Their respective perpendicular angles are $a_\perp = 90^\circ$, $a'_\perp = 135^\circ$, $b_\perp = 112.5^\circ$, and $b'_\perp = 157.5^\circ$.
In this scheme, if we have a perfect entangled state, we should have $|S|=2\sqrt{2}>2$, which breaks the inequality.

For integrated photonics characterization, we use the same circuit as in QST, but with different phase settings. The MZI phase lookup table for CHSH inequality measurement is shown in Fig.~\ref{chshbasis}.

With the 2-fold coincidence counts $cc(a,b)$, we can calculate the correlation function $E(a,b)$ and finally obtain the value of $|S|$. There are 16 angle combinations needed to measure all $cc(a,b)$. In Fig.~\ref{chshbasis}(b), I also provide a scheme using only one pair of detectors.

\section{Integrating entanglement purification and entanglement swapping on silicon chips}

\begin{figure}[h]
    \centering
    \includegraphics[width=12.9cm]{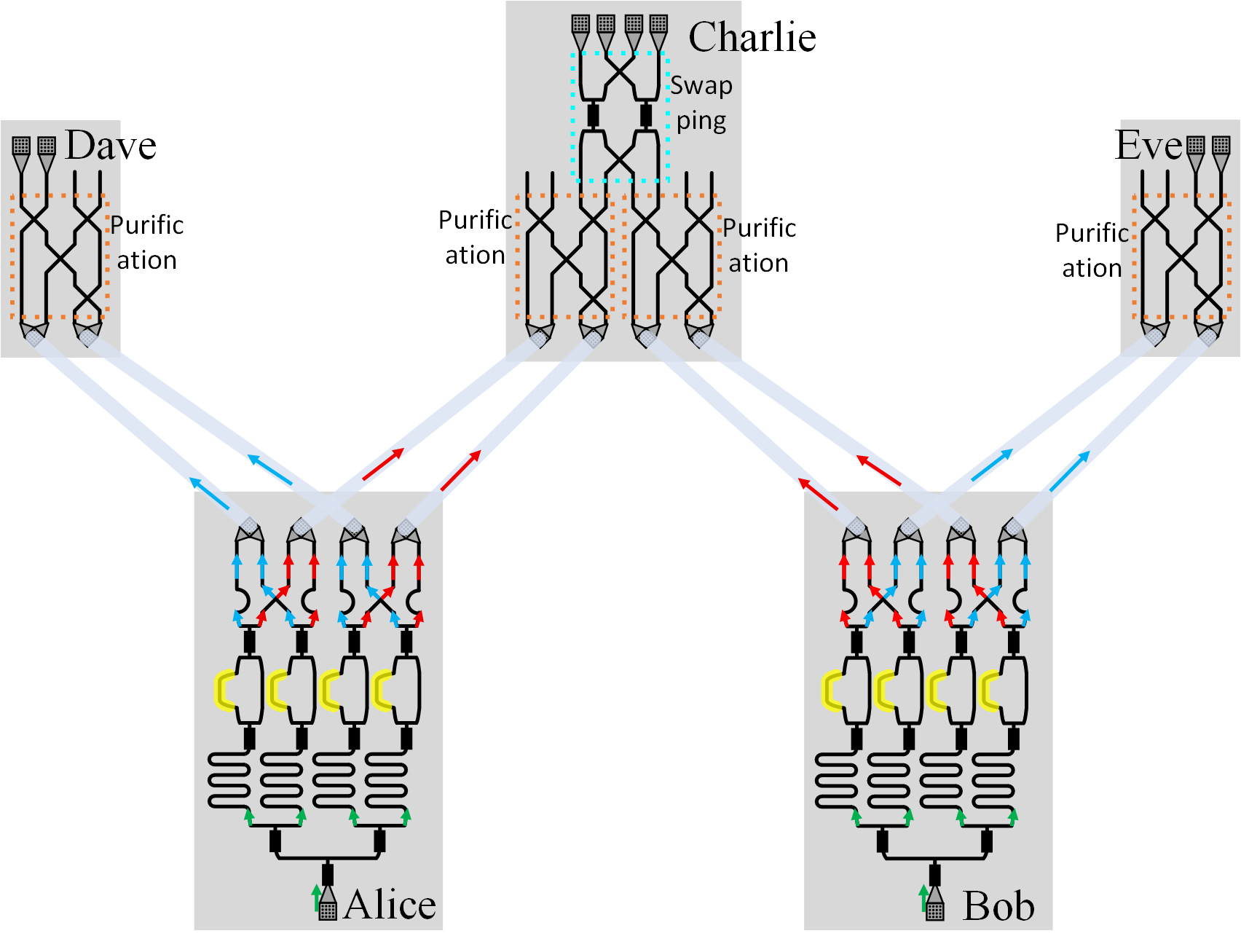}
    \caption{\label{chshbasis} The scheme for integrating entanglement purification and entanglement swapping.}  
\end{figure}

To prove the scalability of our purification circuits, we propose an scheme that integrates entanglement purification and entanglement swapping. The experiment involves five chips, where Alice and Bob serve as hyperentanglement sources, Charlie performs the swapping operation, and Dave and Eve are the two end nodes where entanglement is to be established. Purification is implemented before the measurements of Dave and Eve and prior to the swapping operation at Charlie.

Alice and Bob are pumped to generate entangled photon pairs through SFWM, with 2D GCs converting the high-dimensional entangled state into hyperentangled state $\frac{1}{2}(\ket{00}+\ket{11})(\ket{HH}+\ket{VV})$. For Alice, two pairs of dual fibers are used to transmit the hyperentangled photons: the signal photon is sent to Dave, and the idler photon is sent to Charlie. Similarly, Bob uses one dual fiber to send the idler photon to Charlie and another to send the signal photon to Eve.

Charlie purifies the photons received from Alice and Bob separately using two purification circuits proposed in this work. After consuming the spatial-mode dimension, each photon from Alice or Bob remains in two waveguides, which are then sent to the swapping circuits. The swapping operation is performed using the fusion operator \cite{llewellyn2020chipswapping}. On Charlie’s side, the photons from the four 1D GCs are connected to off-chip SPDs for detection.

Dave and Eve each use purification circuits to purify the photons collected from the 2D GCs. Based on the detection events reported by Charlie, the swapped and purified entanglement between Dave and Eve can be collected from the 1D GCs.

\end{document}